\documentclass[%
 article,
 reprint,
 superscriptaddress,
 amsmath,amssymb,
 aps,
 prd,
]{revtex4-2}
\usepackage{graphicx}
\usepackage{dcolumn}
\usepackage{bm}
\usepackage{siunitx}
\usepackage{mathrsfs}
\usepackage{xcolor}
\usepackage{comment}
\usepackage{adjustbox}
\usepackage[none]{hyphenat}
\usepackage[utf8x]{inputenc}
\usepackage{ucs}
\usepackage{multirow}
\usepackage[normalem]{ulem}
\usepackage{subfigure}
\usepackage{rotating}
\usepackage{ragged2e} 
\usepackage{xparse}

\begin{document}
\title{
SCEP: a Cosmic Magnetic Monopole Search Experiment
}



\def\USTClab{State Key Laboratory of Particle Detection and Electronics, University of Science and Technology of China, Hefei 230026, China}
\def\USTCdep{Department of Modern Physics, University of Science and Technology of China, Hefei 230026, China}
\def\DESL{Deep Space Exploration Laboratory, Hefei, 230022, China}
\def\MagLab{Hefei National Laboratory for Physical Sciences at the Microscale, University of Science and Technology of China, Hefei, 230026, China}
\def\CASLab{CAS Key Laboratory of Microscale Magnetic Resonance, University of Science and Technology of China, Hefei, 230026, China}
\def\SynCenter{Synergetic Innovation Center of Quantum Information and Quantum Physics, University of Science and Technology of China, Hefei, 230026, China}

\affiliation{\USTClab}
\affiliation{\DESL}
\author{Changqing Ye}\thanks{These two authors contributed equally to this work.}\affiliation{\USTClab}\affiliation{\USTCdep}
\author{Beige Liu}\thanks{These two authors contributed equally to this work.}\affiliation{\USTClab}\affiliation{\USTCdep}
\author{Zhe Cao}\email[Corresponding author:]{caozhe@ustc.edu.cn}\affiliation{\USTClab}\affiliation{\DESL}\affiliation{\USTCdep}
\author{Lingzhi Han}\affiliation{\USTClab}\affiliation{\USTCdep}
\author{Xinming Huang}\affiliation{\USTCdep}\affiliation{\MagLab}\affiliation{\CASLab}\affiliation{\SynCenter}
\author{Min Jiang}\affiliation{\USTCdep}\affiliation{\MagLab}\affiliation{\CASLab}\affiliation{\SynCenter}\affiliation{\DESL}
\author{Dong Liu}\affiliation{\DESL}
\author{Qing Lin}\email[Spokesperson \& Corresponding author:]{qinglin@ustc.edu.cn}\affiliation{\USTClab}\affiliation{\DESL}\affiliation{\USTCdep}
\author{Shitian Wan}\affiliation{\USTClab}\affiliation{\USTCdep}
\author{Yusheng Wu}\affiliation{\USTClab}\affiliation{\DESL}\affiliation{\USTCdep}
\author{Lei Zhao}\email[Corresponding author:]{zlei@ustc.edu.cn}\affiliation{\USTClab}\affiliation{\DESL}\affiliation{\USTCdep}
\author{Yue Zhang}\affiliation{\USTClab}\affiliation{\USTCdep}
\author{Xinhua Peng}\email[Corresponding author:]{xhpeng@ustc.edu.cn}\affiliation{\USTCdep}\affiliation{\MagLab}\affiliation{\CASLab}\affiliation{\SynCenter}\affiliation{\DESL}
\author{Zhengguo Zhao}\email[Corresponding author:]{zhaozg@ustc.edu.cn}\affiliation{\USTClab}\affiliation{\DESL}\affiliation{\USTCdep}\

\collaboration{SCEP Collaboration}
\noaffiliation

\date{\today}

\begin{abstract}

Magnetic monopole is a well-motivated class of beyond-Standard-Model particles that could provide insights into the long-standing puzzle of the quantization of electric charge. 
These hypothetical particles are likely to be super heavy ($\sim$10$^{15}$\,GeV) and  be produced in the very early stages of the Universe's evolution. We propose a novel detection scenario for the search of such cosmic magnetic monopoles, utilizing a hybrid approach that combines radio-frequency atomic magnetometers and plastic scintillators. Such setup allows for the collection of both the induction and scintillation signals generated by the passage of a magnetic monopole, which provides acceptance to the magnetic monopoles with their velocities larger than about 10$^{-6}$ light speed (assuming a signal-to-noise ratio of $\sim$4) and their masses larger than approximately 10$^7$\,GeV (at $\beta\sim10^{-3}$). 
The proposed detector design has the potential to scale up to large area, enabling the exploration of the parameter space of the cosmic magnetic monopole beyond the current experimental and astrophysical constraints. It is estimated that such detector can reach current most stringent limits of the flux  set by previous searches, with a signal-to-noise ratio of the induction signal larger than about 4.5, assuming an effective exposure being 20000\,year$\cdot$m$^2$ and coil layer of 3.

\end{abstract}

\maketitle

\newcommand{\mwba}[1]{\textcolor{violet}{#1}}
\newcommand{\mwbd}[1]{\textcolor{violet}{\sout{#1}}}

\section{Introduction}


A magnetic monopole (MM), proposed by Dirac in 1931~\cite{dirac1931quantised}, is a theoretical particle postulated to exist as an isolated source of a singular magnetic charge, analogous to the electric charge. 
The MMs hold significance in fundamental physics as they provide a means to  explain the quantization of electric charge. 
Dirac derived that the minimal magnetic charge of MM is $g_m = 4.14125 \times 10^{-15}$ Wb.
The MM with this unit magnetic charge (denoted as Dirac MM) is the benchmark for most of the MM searches.
The concept of the MMs also finds natural incorporation within the framework of Grand Unified Theories (GUTs)~\cite{georgi1974unity}, which aim to unify the electromagnetic, weak, and strong nuclear forces. 
The quantization of electric charge is also explained in the framework of GUT.
Most GUT models generally predict that large numbers of the MMs are created when the strong force decouples from the electro-weak force.
The typical energy scale is $\sim$10$^{15}$\,GeV which also determines the mass scale of the GUT-MMs.
The observed lack of the GUT-MMs in the universe today is one of the key motivations behind the proposal of the cosmological inflation~\cite{zeldovich1978concentration, guth1981inflationary}.

The searches for these cosmic MMs predicted by the GUT models persist through various experimental approaches, including the ultra-low background experiments and the superconducting coil-based experiments.
More information about the searches of the MMs can be found in Ref.~\cite{preskill1984magnetic, patrizii2015status}.
The ultra-low background experiments are typically conducted in the underground environments with kilometers of rock overburdens, providing a shielding against the cosmic-ray backgrounds. 
These experiments aim to detect the ionization or scintillation signals produced by the MMs as they traverse the target material of the detector.  
Notably, the MACRO experiment~\cite{macro2002final} and the neutrino telescope IceCube~\cite{abbasi2013search} have yielded the most sensitive searches for the MMs with speeds greater than about 4×10$^{-5}$ times the light speed and with relativistic speeds, respectively. 
While the ionization density caused by the MMs is predicted to be higher (particulary for high-speed MMs) than that of the background particles commonly observed in the terrestrial detectors, such as muons and electrons, due to the unit magnetic charge being much larger than the unit electric one.
It is important to consider the possibility of alternative exotic particles, such as the superheavy dark matter~\cite{bertone2005particle, chung1998superheavy}, which could also create high-density ionization. 
Conversely, superconducting coil-based experiments~\cite{eberhard1975improvements, cabrera1982first} focus on detecting the smoking-gun induction signals generated by the MMs. 
However, these experiments face limitations in terms of their sizes due to the requirements of maintaining superconducting temperatures.
It is also worth noting that MMs predicted by non-GUT models can have distinct masses, velocities, and magnetic charges (see Ref.~\cite{preskill1984magnetic, rossi1982exact, weinberg2007magnetic, mavromatos2020magnetic} for theoretical reviews). 
Dirac MMs with magnetic charge of $g_m$ is usually set as primary search target to be a conservative approach. 
Numerous scientific experiments have been conducted to search for such non-GUT MMs.  
For instance, experiments performed near particle colliders have been utilized to investigate the potential existence of the low-mass MMs~\cite{acharya2014physics, bertani1990search, abbiendi2008search, kinoshita1992search, pinfold1993search}. 
Additionally, experiments that collect and analyze the terrestrial and extraterrestrial samples~\cite{price1988limits, ghosh1990supermassive, bendtz2013search} have been employed to search for the MMs that may be bound to matter.


This article presents a comprehensive illustration of the SCEP (Search for Cosmic Exotic Particles) experiment, with a specific emphasis on the detection perspective of mainly the cosmic MMs.
We propose a novel approach utilizing a coincidence measurement technique that uses the radio-frequency atomic magnetometers and the plastic scintillators (PSs). 
The primary components of this MM detector are room-temperature induction coils and PSs, which make it a cost-effective and scalable system.
Simulations have shown that this detection setup is sensitive to MMs with velocities $\beta > \sim10^{-6}$ (assuming a signal-to-noise ratio of about 4)~\footnote{Further increasing the signal-to-noise ratio of the induction detection can extend the search to lower velocity region.}.
Such searches for the cosmic MMs can be carried out at the sea-level or high-altitude environment, rather than in an underground laboratory, taking advantage of the reduction in the charged particle background achieved through the coincidence requirement between the induction coils and PS detectors.
A surface or high-altitude detector deployment can lower the mass threshold of the cosmic MM search from about 10$^{10}$\,GeV to 10$^7$\,GeV~\cite{patrizii2019searches} (at $\beta\sim10^{-3}$) compared to an underground detector, due to the reduced overburden.
This will be detailed also in Sub-section~\ref{subsec:earth_shielding}.
Based on these advantages, this detection scheme holds potential for the large-scale deployment, surpassing the current best flux limits~\cite{macro2002final, abbasi2013search, giacomelli1984magnetic} and the astrophysical constraints~\cite{adams1993extension, lewis2000protogalactic} for the cosmic MMs.

The fundamental concept of the detector system is illustrated in Section~\ref{sec:detector_concept}.
For estimating the sensitivities to the cosmic MMs using the proposed system, we have developed a  simulation framework of the induction signal, which is  described in Section~\ref{sec:simulation_of_signal}.
The validation of the simulation is performed and described in Section~\ref{sec:validation}.
Furthermore, the estimation of the background and sensitivity of the SCEP experiment to the cosmic MMs are presented in Section~\ref{sec:sensi_to_mm}.

\section{Detector Concept}
\label{sec:detector_concept}

\begin{figure}[htp]
    \centering
    \includegraphics[width=0.9\columnwidth]{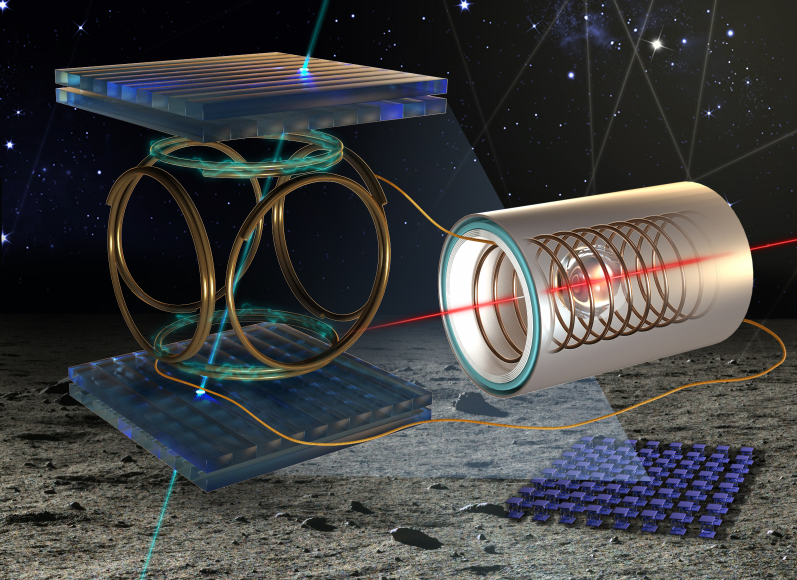}
    \caption{Schematic diagram of single module detector of SCEP.}
    \label{fig:scep_schematic_diagram}
\end{figure}

A single module of the SCEP detector encompasses dedicated the detection systems for both the scintillation and induction signals, as illustrated in Fig.~\ref{fig:scep_schematic_diagram}. 
The scintillation signals are captured by the PSs positioned at the top and bottom of the module. 
In the preliminary design, each PS module is constructed using the designs similar to the ones utilized in ~\cite{zhou2020time}.
To guide the scintillation light, wavelength-shifting fibers are incorporated within the PS module. 
These fibers serve the purpose of directing the emitted light to Silicon photomultipliers (SiPMs) coupled at the ends of the fibers.
A preliminary simulation using the GEANT4 toolkit~\cite{agostinelli2003geant4} has been conducted to evaluate the performance of the PS module. 
The obtained results indicate a light yield of approximately 22 photoelectrons (PE) per MeV, thereby enabling an energy resolution of about 8.6\% and 2.5\% for muons at $\sim$8\,MeV and Dirac MMs at $\sim$ 100\,MeV, respectively.

\begin{figure*}[htp]
    \includegraphics[width=0.97\textwidth]{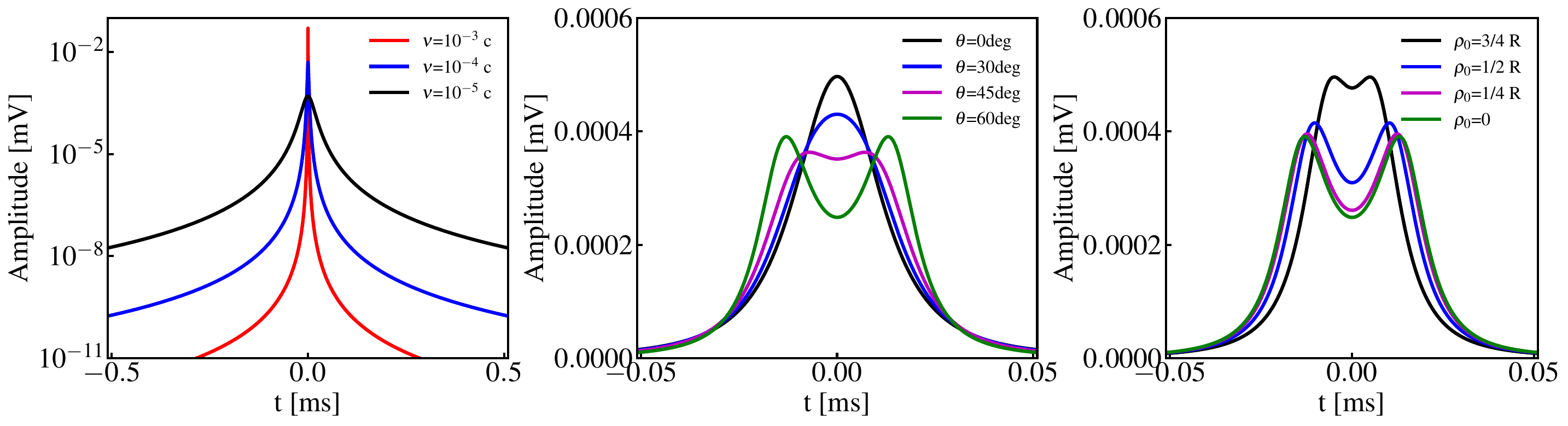}
    \caption{
    Simulated waveforms for the induction voltages by MMs.
    The left, middle, and right panels show the induction signals with different MM velocities, polar angles, and transverse distance to the coil center respectively.
    The waveforms in the middle and right panels are calculated assuming $\nu$=10$^{-5}$c.
    These waveforms are calculated assuming an induction coil with 12-cm diameter and about 4320 turns.
    }
    \label{fig:mm_waveforms}
\end{figure*}

The induction signals resulted from the passage of a MM are collected using an apparatus that integrates an induction coil, a Helmholtz coil, and a magnetometer.
The micro-current induced by the MM passing through the induction coil is subsequently directed to the Helmholtz coil, leading to the generation of an alternating magnetic field at the center of the Helmholtz coil, which is wound around the core gas chamber of a radio-frequency atomic magnetometer.
Subsequently, the alternating magnetic field is detected utilizing this kind of magnetometer renowned for its exceptional sensitivity to magnetic field.
In our detector setup, a magnetometer with a sensitivity of several fT/$\sqrt{\textrm{Hz}}$ to the magnetic field will suffice since the dominant noise comes from the thermal noise of the induction coil.
The target material (for example, rubidium 87) of the magnetometer is confined in a transparent gas chamber, and is heated to maintain a temperature of around 200 degrees Celsius.
The material is also polarized by a static magnetic field aligned along the $Z$ axis with the assistance of a beam of pump laser.
The presence of an alternating magnetic field in the $XY$ plane can impact the precession of the atoms within the gas chamber.
This effect manifests as the variations in the measured light intensity of a laser beam which pass through the gas chamber.
More details of the working principles of the  magnetometer are given in Ref.~\cite{lee2006subfemtotesla, savukov2005tunable, savukov2007detection, jiang2020interference, jiang2021search}.
A preliminary prototype of the magnetometer can reach a detection sensitivity of $<$10\,fT/$\sqrt{\textrm{Hz}}$ for the alternating magnetic field~\cite{kominis2003subfemtotesla, hong2024femtotesla}.
The frequency range of the magnetometer readout is approximately from several hundred Hz to several tens of kHz, limited by the Larmor precession frequency of the target atoms in the magnetometer under the static polarization magnetic field that can be practically applied.
Additionally, we consider an alternative readout configuration. 
In this configuration, the induction coil is directly connected to an operational amplifier (OPA) and an analog-to-digital converter (ADC). 
Although this configuration has higher intrinsic back-end noise levels compared to a magnetometer readout, it can reach a higher frequency range up to the MHz level, potential extending the induction sensitivity for the fast MMs. 
This configuration also has other advantages, such as a more compact size conducive to integration, a more cost-effective design, and reduced overall weight.


\section{Simulation of Signal}
\label{sec:simulation_of_signal}
The signal-to-noise ratio (SNR) to a single Dirac MM stands as a critical measure of the quality for the induction signal in this work. 
The SNR in this work is defined as the maximum signal amplitude squared $A^2_S$ divided by the mean-squared noise amplitude $\langle A^2_N \rangle$:
\begin{equation}
    \textrm{SNR} = \frac{ \textrm{MAX} (A^2_S)}{\langle A^2_N \rangle}.
\end{equation}
Larger values of SNR are preferred for higher noise rejection power.
The SNR of the system is related to various factors, most dominantly the resistance of the induction coil, which introduces significant thermal noise. 
Additionally, the prevailing temperature conditions, as well as the signal response characteristics of the circuit and the magnetometer, also plays a role.
To estimate the SNRs for MMs with various velocities, a comprehensive simulation framework has been developed which is described briefly in the following subsections.

\begin{figure*}[htp]
\begin{equation*}
\scriptsize
    U = -\frac{g_mn}{2\pi} \frac{(z_N^2 + (1 - \rho_N)^2)(z_N \rho_T - \rho_N z_T)  \mathcal K \left(\frac{4 \rho_N}{z_N^2 + (1 + \rho_N)^2}\right) + (z_T \rho_N (\rho_N^2 + z_N^2 - 1) - z_N (1 + z_N^2 + \rho_N^2) \rho_T) \mathcal E \left(\frac{4 \rho_N}{z_N^2 + (1 + \rho_N)^2}\right)}{(z_N^2 + (1 - \rho_N)^2)  \rho_N \sqrt{z_N^2 + (1 + \rho_N)^2}},
\end{equation*}
\hfill
\begin{equation}
\textrm{with}
\left\{
    \begin{aligned}
    & \rho_N  =  \sqrt{ (v_N  t   \sin\theta)^2 + (\rho_0/R)^2 + 2   v_N   t  (\rho_0/R)   \sin\theta   \cos\phi } &\\ 
    & \rho_T = (v_N (\rho_0/R)  \cos\phi \sin\theta + v_N^2  t \sin^2\theta)/\rho_N & \\
    & z_N  =  v_N   t   \cos\theta & \\
    & z_T = v_N \cos\theta & \\
    & v_N = v/R & \\
    \end{aligned}
\label{eq:mm}
\right.
\end{equation}
\end{figure*}

\subsection{Induction}

The induction voltage on the induction coil is calculated assuming that the thickness of the coil brings negligible effect.
When a MM with the velocity of $v$ passing through the induction coil which has a radius of $R$, the induction voltage $U$ can be written as in Eq.~\ref{eq:mm}.
Eq.~\ref{eq:mm} is based on the assumption that the time $t$ is 0 when the MM crosses the coil plane ($z=0$).
$\rho_0$ is the transverse distance to the coil center when the MM reaches the coil plane. 
$\theta$ and $\phi$ represent the polar and azimuth angles, respectively, of the incoming MM's direction under the spherical coordinates with the $z$ axis perpendicular to the coil plane.
$n$ is the coil turn number, the $\mathcal K$ and $\mathcal E$ functions are the complete elliptic integrals of the first and second kinds.
The induction signal is at maximal when the MM passes through the coil center with $\theta=0$.
The amplitude and spectral shape of the induction signals are predominantly influenced by the MM speed, the polar angle, and the transverse distance to the coil center.  
These dependencies are shown in Fig.~\ref{fig:mm_waveforms}.


The interaction of charged Standard Model (SM) particles, or SM particles with magnetic moments, with the induction coils can potentially result in induction signals.
However, there are significant distinctions in the amplitude and spectral shape of these induction signals compared to those induced by the MMs.
More importantly, the induction signals generated by the SM particles have a vanishing time integral due to their nature as, at most, magnetic dipoles.
In addition, the common background SM particles, such as the muons, neutrons, and protons, typically exhibit relativistic speeds, leading to the rapid resonant induction on the timescale of approximately 10 picoseconds for a coil with a 12-centimeter diameter.
The quick oscillation of the voltage cannot be effectively shaped by the subsequent relatively ``slow'' circuitry and reliably detected by the read-out devices.
Considering these factors, the induction caused by SM particles is considered to be negligible.

\subsection{Signal shaping}

\begin{figure}
    \centering
    \includegraphics[width=0.9\columnwidth]{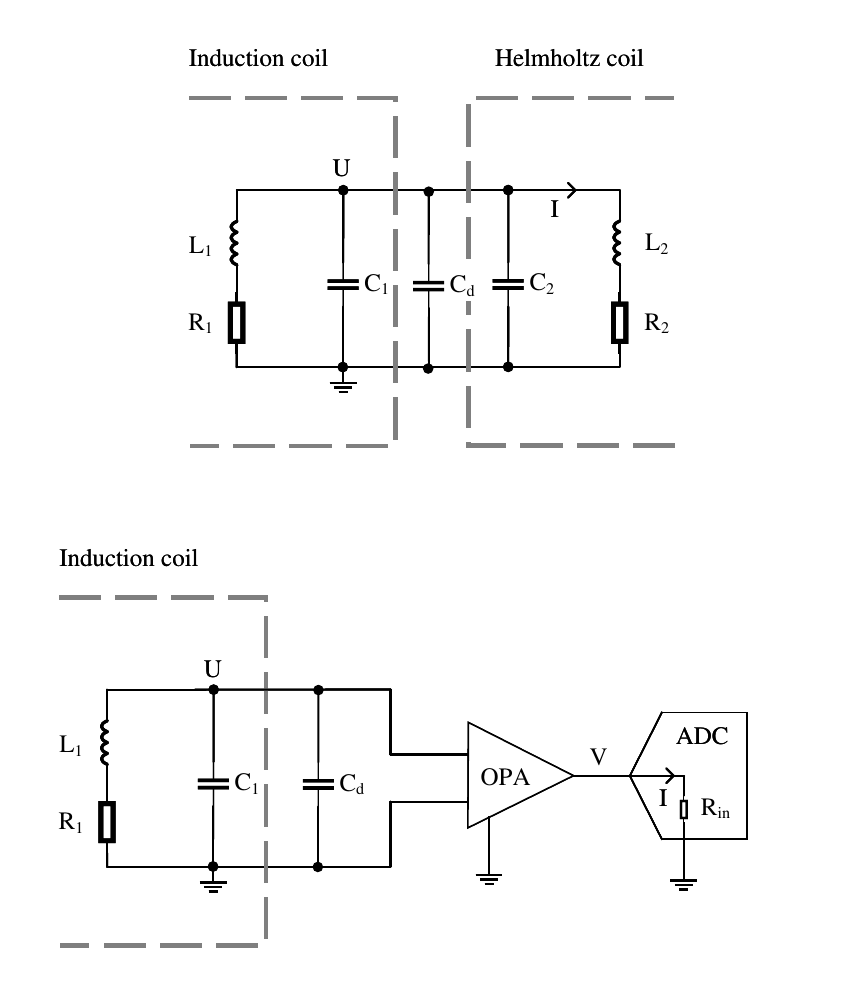}
    \caption{
    Circuit diagrams for the magnetometer readout scenarios (top) and ADC readout (bottom). }
    \label{fig:coil_circuit_model}
\end{figure}
The induction coils possess non-trivial resistances, capacitances, and inductances, which affect both the amplitude and temporal characteristics of the electric current within the circuit.
In the signal simulation,  it is assumed that a single induction coil can be approximated as a series combination of a resistor and an inductor, paralleled by a capacitor.
The circuit diagram of the induction and Helmholtz coils in the magnetometer-readout scenario, as well as of the direct readout scenario using the ADC, is depicted in Fig.~\ref{fig:coil_circuit_model}.
In the circuit diagram, $L_1$ ($L_2$), $R_1$ ($R_2$), and $C_1$ ($C_2$) are the effective inductance, resistance, and capacitance, respectively, of the induction (Helmholtz) coil.
$C_d$ represents other parallel capacitive components in the circuit, mainly the distributed capacitance of the cable and the input capacitance of the OPA.
$U$ is the induction voltage, and $I$ is the induction current on the Helmholtz coil which is directly related the strength of magnetic field that is eventually captured by the magnetometer in the magnetometer-readout scenario.
In the alternative direct ADC-readout scenario, $V$ represents the voltage detected by the ADC, while $I$=$V/R_{in}$ represents the electric current flowing into the ADC. $R_{in}$ is the coupling resistance of the ADC.

\begin{figure}
    \centering
    \includegraphics[width=0.9\columnwidth]{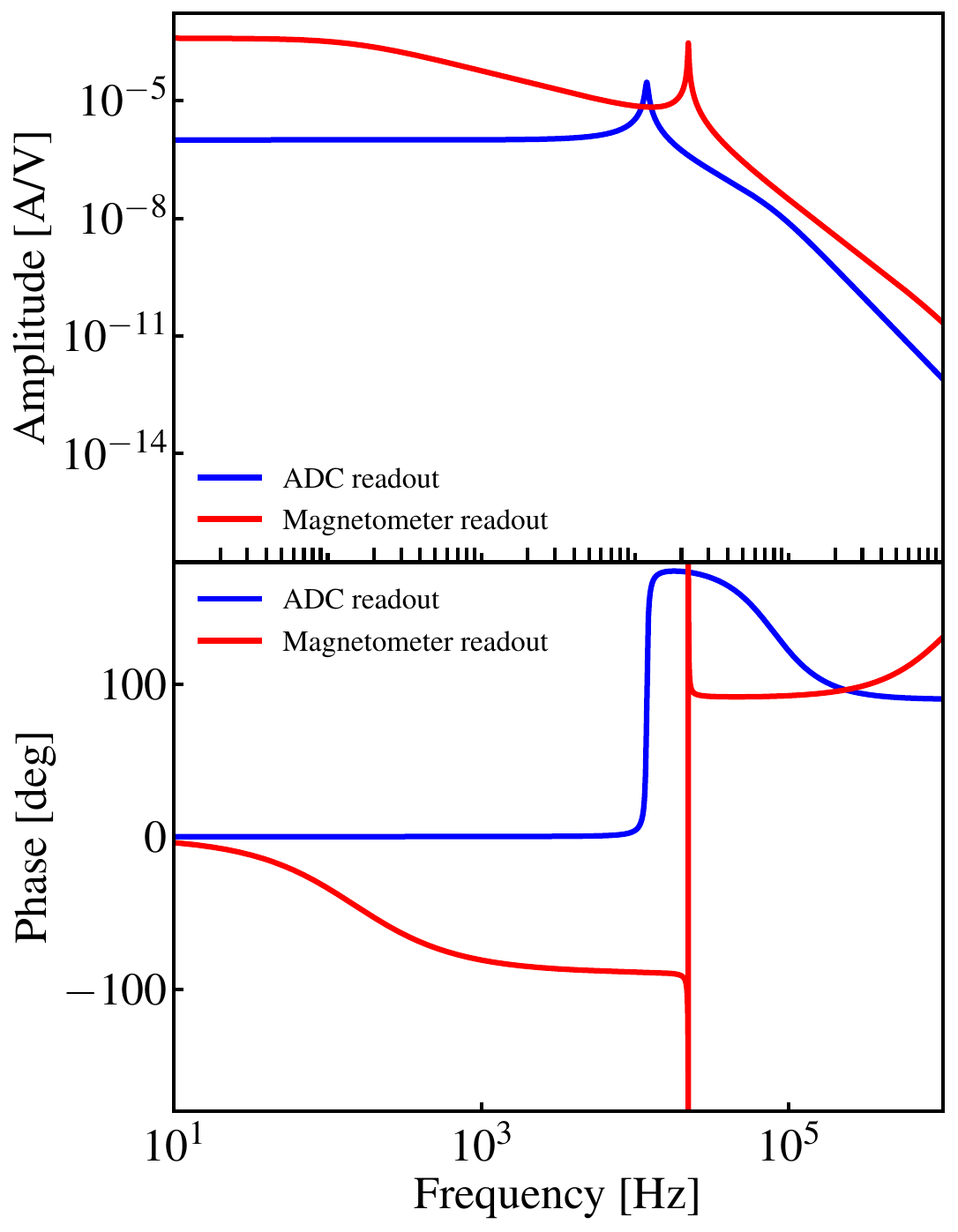}
    \caption{
    Response functions of the circuits with different readout scenarios.
    The top and bottom panels show the amplitude and phase spectra, respectively.
    The red and blue lines represent the ADC and magnetometer readout scenarios.
    }
    \label{fig:response_function}
\end{figure}

\begin{table*}[htp]
    \footnotesize
    \begin{tabular}{c|c|c|c|c|c|c|c}
    \hline\hline
    \multirow{2}{*}{Coil}                               & 
    \multirow{2}{*}{Wire type}                          &
    \multirow{2}{*}{Wire diameter [mm]}                 &
    \multirow{2}{*}{Minimal coil radius [cm]}           &
    \multirow{2}{*}{Maximal coil radius [cm]}           &
    \multirow{2}{*}{Turn number}                        &
    \multicolumn{2}{c}{SNR$_0$}   \\
    \cline{7-8}
     & & & & & & ADC readout & Mag. readout \\
    \hline\hline
    V1                                  &
    Simple                              &
    0.11                                &
    5.7                                 &
    7.2                                 &
    4320                                 &
    0.16                                 &
    0.16                                 \\
    V2                                  &
    Litz                                &
    1.35                                &
    5.7                                 &
    7.2                                 &
    720                                &
    0.02                                 &
    1.92                                 \\
    V3                                  &
    Simple                              &
    0.55                                &
    10.0                                &
    14.5                                &
    12500                               &
    0.57                                 &
    0.82                                 \\
    \hline\hline
    \end{tabular}
    \caption{
    The geometrical parameters of three test induction coils.
    The last two columns show the best SNRs that can be reached for each test coil by optimizing the circuit configuration for the ADC and magnetometer readout scenarios.
    The SNRs listed are based on the assumption that the MM perpendicularly crosses the coil center with $\beta=10^{-5}$.
    }
    \label{tab:coils}
\end{table*}

\begin{figure}
    \centering
    \includegraphics[width=0.9\columnwidth]{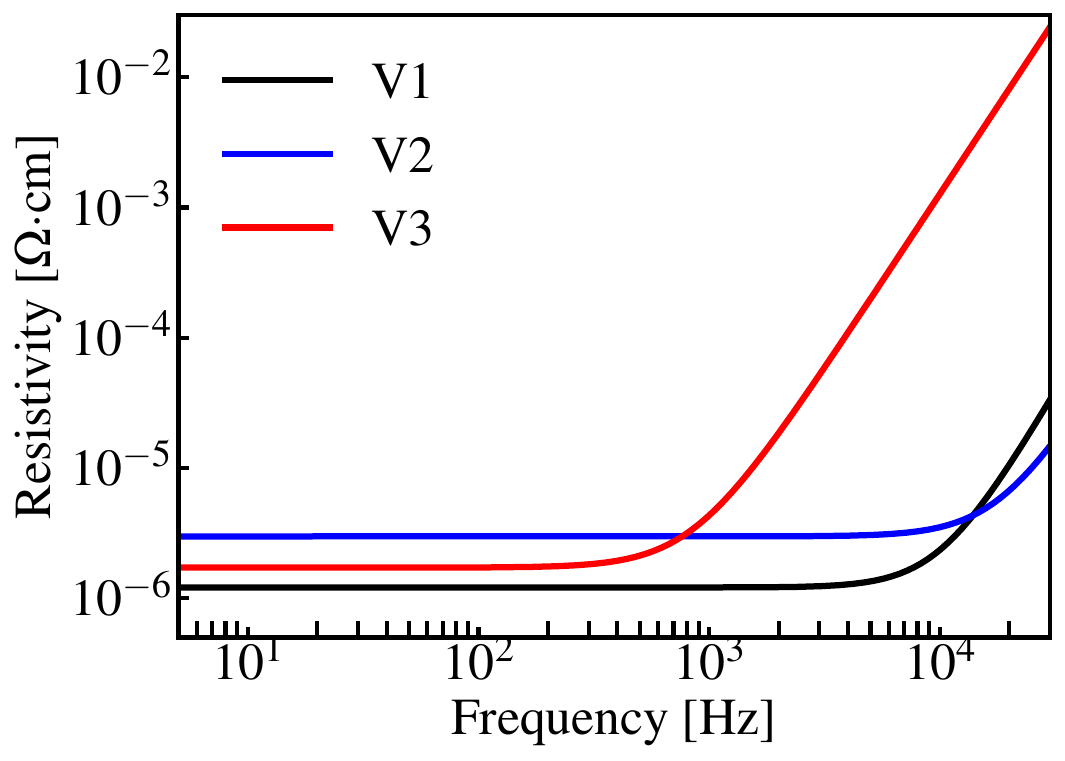}
    \caption{
    Measured frequency-dependent resistivities as a function of signal frequency for the test induction coils. 
    }
    \label{fig:lcr_measurement}
\end{figure}

\begin{figure}
    \centering
    \includegraphics[width=0.9\columnwidth]{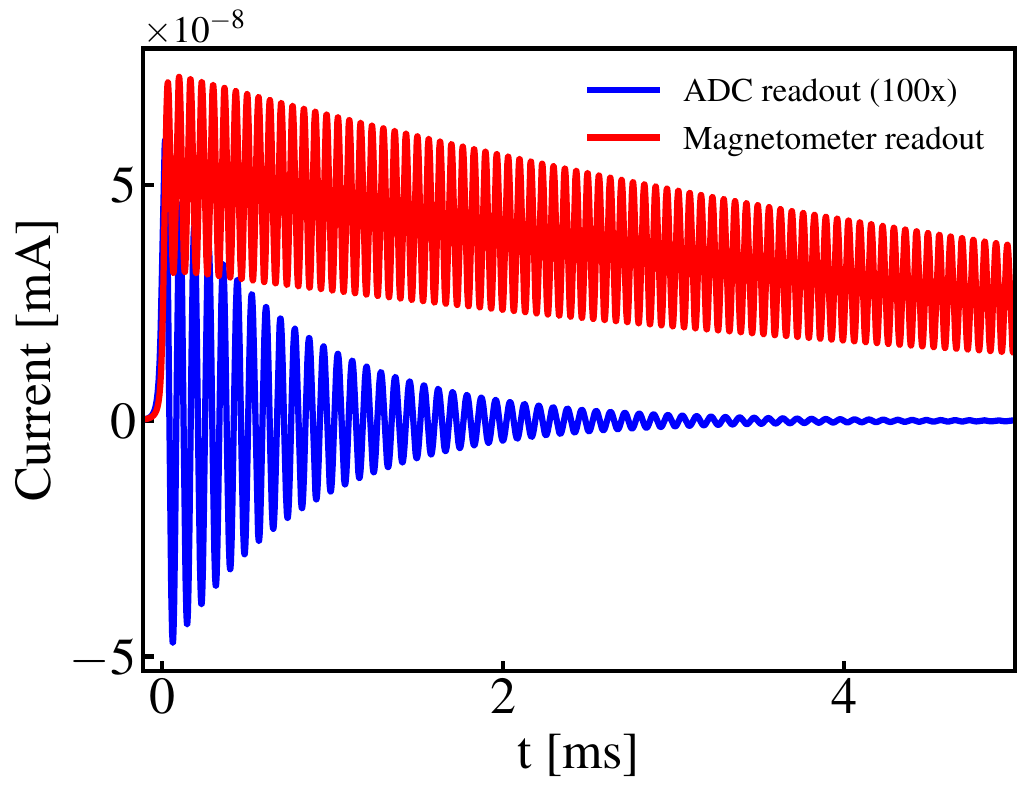}
    \caption{
    The waveforms of the electric currents after the circuit shaping.
        The red and blue solid lines represent the V1 coil in the magnetometer and ADC readout scenarios, respectively.
        The MM speed is assumed to be $\beta$=10$^{-5}$.
    To increase the visibility, the current from ADC readout is amplified by 100 times.
    }
    \label{fig:coil_current}
\end{figure}

In the signal simulation, the electric current $I$ is determined by applying a circuit response function to the induction voltage $U$. 
The Fourier transform of the electric current, denoted as $i(\omega)$, can be expressed as:
\begin{equation}
    i(\omega) = u(\omega) \cdot \mathcal H(\omega)
\end{equation}
where the complex $u(\omega)$ is the Fourier transform of the induction voltage.
The circuit response function $\mathcal H(\omega)$ is analytically derived based on the effective circuit models shown in Fig.~\ref{fig:coil_circuit_model}.
A response function for a 6-cm-radius coil with 4320 turns is presented in Fig.~\ref{fig:response_function}. 
The most sensitive frequency range (frequencies around the resonant peaks in Fig.~\ref{fig:response_function}) is determined mostly by the circuit configuration. 
The resonant frequency for the magnetometer-readout scenario needs to be adjusted so that it matches the intrinsic resonant frequency of the Larmor precession frequency of the target gas.
The resonant frequencies of the two readout scenarios exhibit variations owing to disparities in the circuit configurations.
In particular, the inductance $L_2$ and capacitance $C_2$ of the Helmholtz coil in the magnetometer readout scenario contribute to a higher resonant frequency compared to the alternative ADC readout scenario.
Among the various electric parameters, the resistance of the induction coil $R_1$ is identified as most dominant. 
This resistance depends on the frequency $\omega$, mainly due to the presence of the skin effect and the proximity effect~\cite{dowell1966effects}.
However, the exact relation between the coil resistance and signal frequency cannot be analytically given due to the complexity of the coil structure.
To investigate the frequency dependence of the coil resistance, \textit{in-situ} measurements are conducted using an HIOKI IM3533-01 LCR meter. 
The magnitude is denoted as $Z_c$, and the phase angle is denoted as $\theta_c$.
The $Z_c$ and $\theta_c$ have correlation with the inductance $L$, capacitance $C$, and resistance $R_{AC}$ of the coil, which can be expressed as:
\begin{equation}
    \begin{aligned}
        Z_c & =  \sqrt{\frac{{R_{AC}}^2+{\omega}^2L^2}{1-2\omega^2LC+\omega^2C^2({R_{AC}^2}+\omega^2L^2)}} \\
        \theta_c & =  \frac{\omega(L-C{R_{AC}}^2-\omega^2L^2C)}{R_{AC}}.
    \end{aligned}
\end{equation}
The alternating resistance of the induction coil $R_{AC}$ is empirically parameterized as ~\cite{somalwar1988transient}: 
\begin{equation}
 R_{AC} (\omega)=\kappa \omega^{\zeta}+R_{DC},
\end{equation}
where $R_{DC}$ is the frequency-independent direct resistance of the coil.
The parameters $\kappa$ and $\zeta$ are empirical model parameters.
A Nelder-Mead fitting algorithm is utilized to derive the $R_{DC}$, $\kappa$, and $\zeta$ from the set of measured $Z_c$ and $\theta_c$.
In a primary test, three induction coils with different radii and turn numbers are manufactured and tested.
Their geometric parameters are given in Table~\ref{tab:coils}. 
The derived $R_{\textrm{AC}}$ for these induction coils (normalized to resistivity) are shown in Fig.~\ref{fig:lcr_measurement}.
The waveforms of the induction electric current on the Helmholtz coil in magnetometer readout scenario and of electric current flowing into ADC in alternative direct readout scenario (the current $I$ in Fig.~\ref{fig:coil_circuit_model}) are shown in Fig.~\ref{fig:coil_current}, assuming the MM perpendicularly traverses the coil center of V1 with a speed of 10$^{-5}$ light speed.

\subsection{Detection}

\begin{figure*}[htp]
    \centering
    \includegraphics[width=0.9\columnwidth]{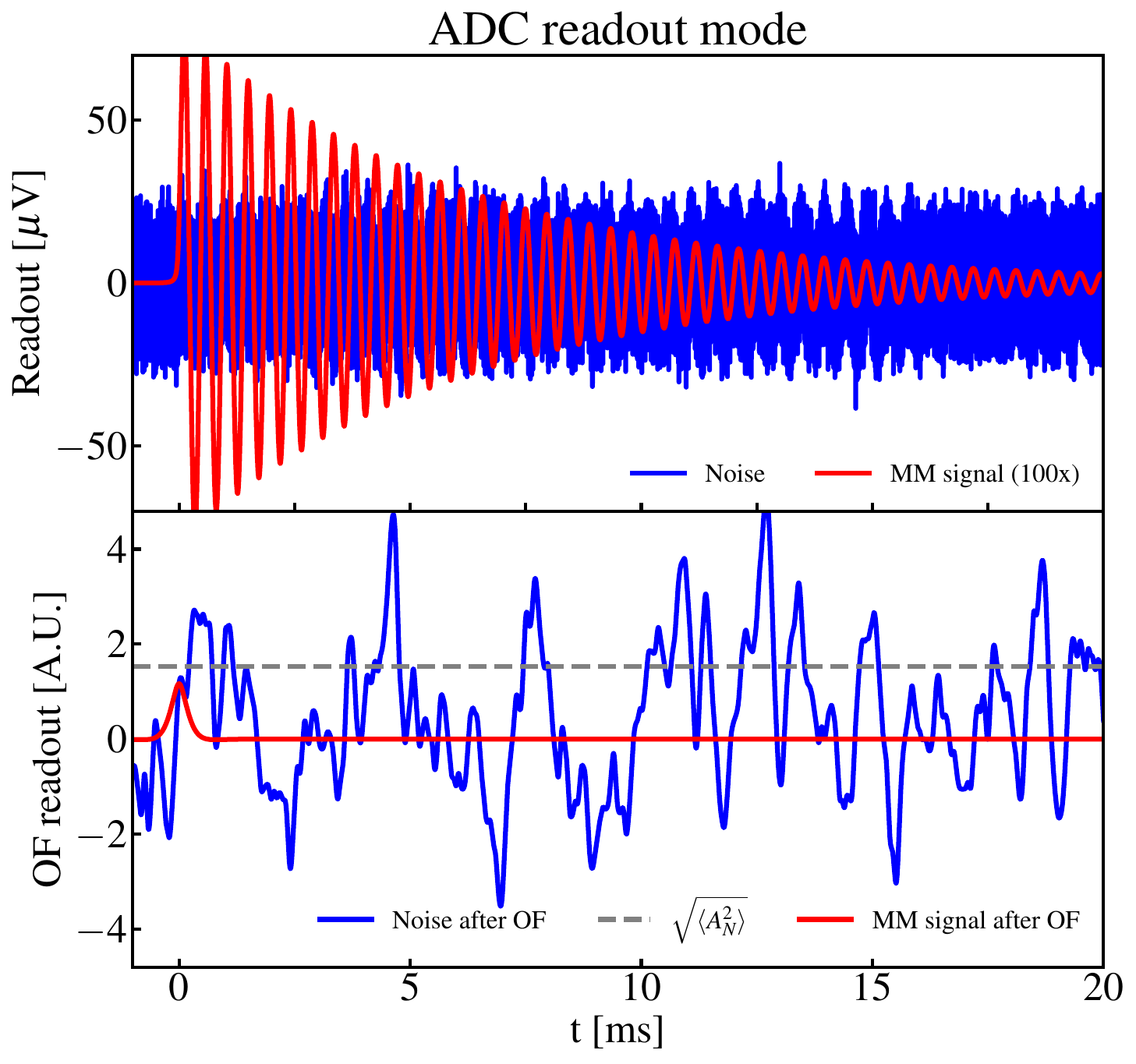}
    \includegraphics[width=0.9\columnwidth]{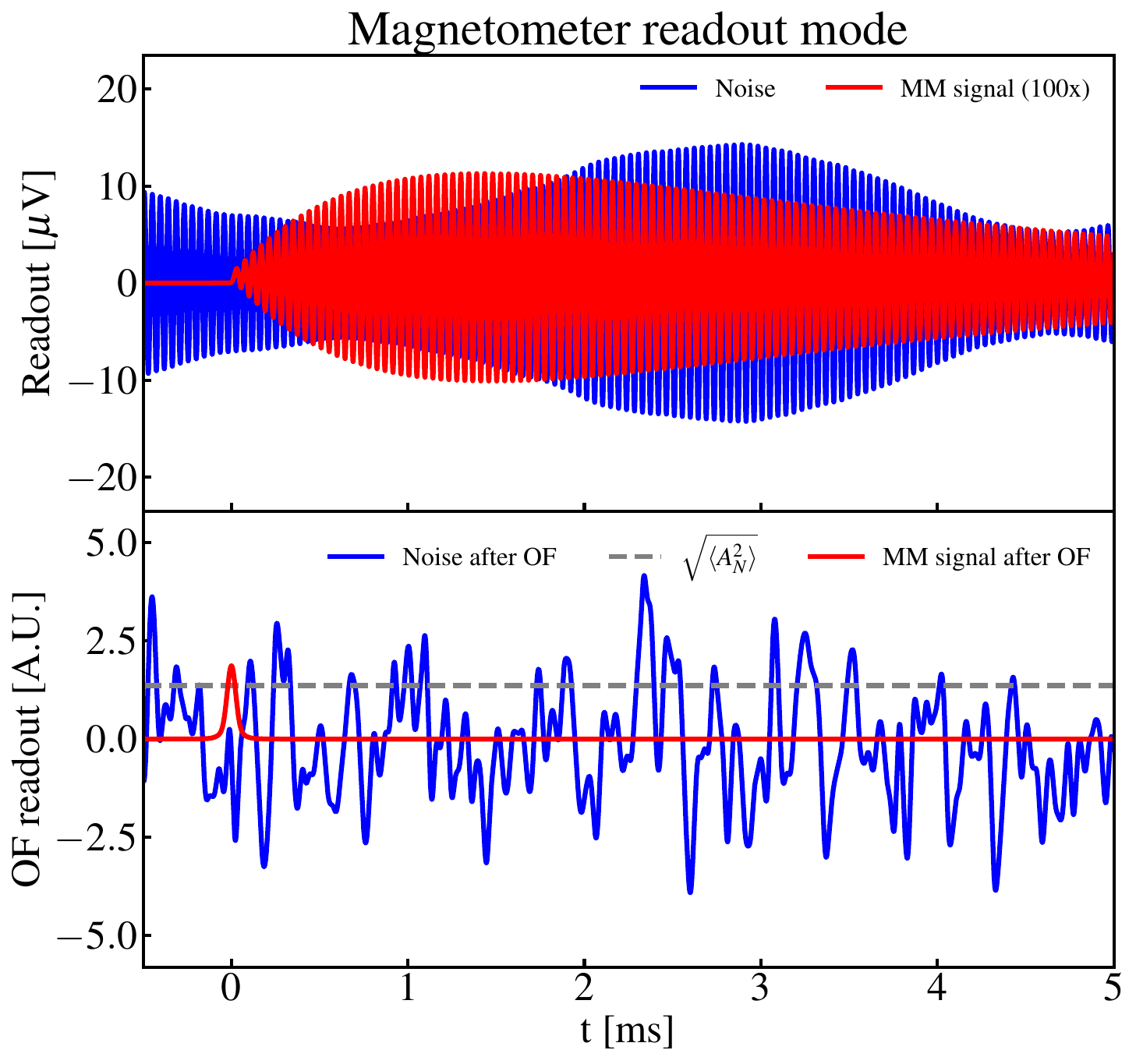}
    \caption{
    The waveforms of readout signals for the ADC and magnetometer readout scenarios are displayed in the left and right panels, respectively.
    The top and bottom panels show the waveforms before and after OF applied.
    The V3 and V2 coils are used for the ADC and magnetometer readout scenarios, respectively, having an average SNR$_0$ to the MM of 0.57 and 1.92.
    The red and blue solid lines correspond to the waveforms of MM signal and noise.
    To enhance visibility, the MM signal waveforms before the OF applied are scaled by 100 times.
    The dashed gray lines give the root-mean-square of the noise amplitude.
    The MM speed is assumed to be $\beta$=10$^{-5}$.
    }
    \label{fig:before_after_of}
\end{figure*}

\begin{figure}[htp]
    \centering
    \includegraphics[width=0.9\columnwidth]{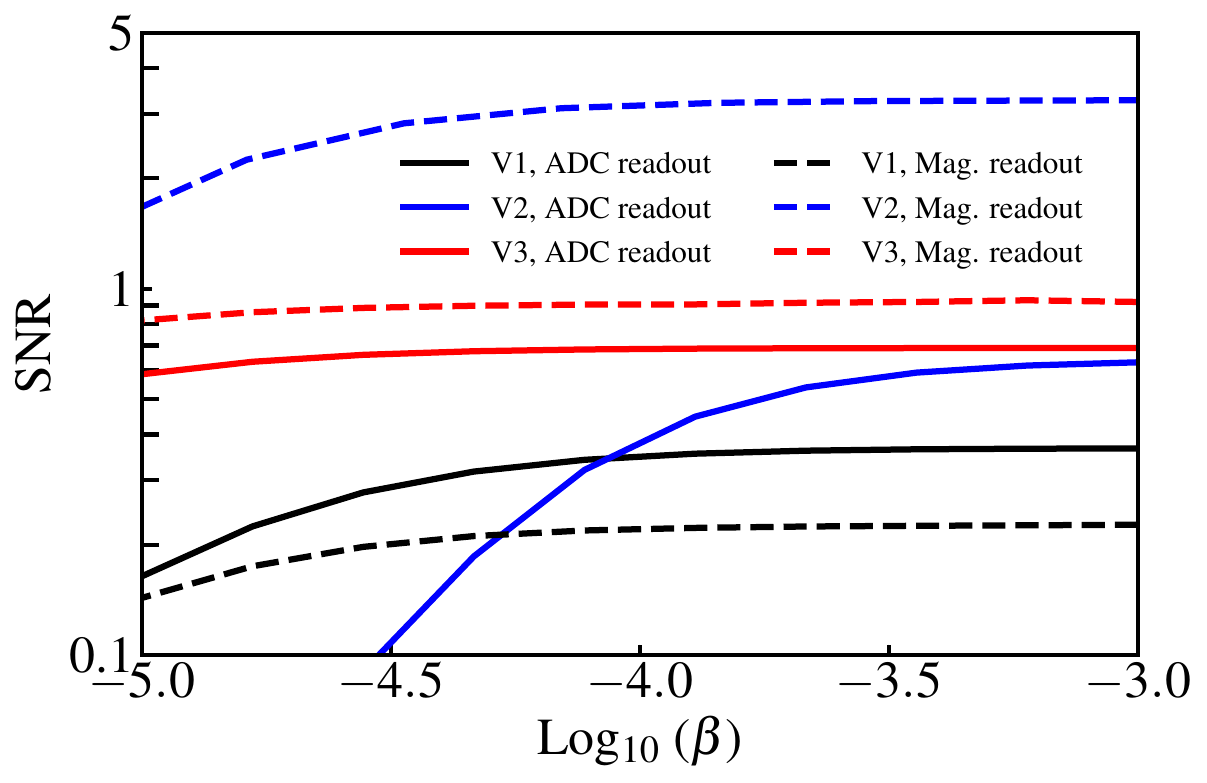}
    \caption{
    The SNR as a function of MM speed for the prototype induction coils for the ADC-readout and magnetometer-readout modes.
    }
    \label{fig:snr_dependence_on_speed}
\end{figure}

The readout devices exhibit diverse response characteristics to the induction signals, owing to their distinct intrinsic mechanisms. 
The magnetometer relies on the atomic precession and typically demonstrates a response timescale ranging from several tens of microseconds to several milliseconds.
The complex response function of the magnetometer $\mathcal H_m$ is commonly modeled in the form of a Lorentzian distribution:
\begin{equation}
    \mathcal H_m (\omega) = \frac{\gamma T_2}{2j+2T_2(\omega_0 -\omega)},
\label{eq:magnetometer_response}
\end{equation}
where $\gamma = \frac{2\mu_B}{5\hbar}$, with $\mu_B$ being the Bohr magnetic moment and $\hbar$ the reduced Planck constant, is the gyromagnetic ratio of the target atom and $T_2$ is the spin relaxation time which can be measured experimentally.
The mean resonant frequency $\omega_0$ of the Lorentzian response function of the magnetometer depends on the applied static magnetic field along the $z$ axis within the gas chamber. 
In the case of the ADC direct readout, the response function can be simplified and approximated as a constant which is dependent on the gain of the OPA and the input impedance of the ADC, within the bandwidth of interest.

\subsection{Reconstruction and thermal noise}

\label{subsec:recon_and_thermal_noise}

To extract the MM signals from a significant amount of noise, the readout output undergoes signal filtering to obtain the final signals. 
In our case, the optimal filter (OF) method is applied for signal extraction.
The response function of OF, denoted as $\mathcal H_{\textrm{OF}}$, can be written as~\cite{somalwar1988transient}:
\begin{equation}
\mathcal H_{\textrm{OF}} = \frac{u^\ast (\omega)}{S_n(\omega) \prod \mathcal H_i^{\ast} (\omega)+S_H(\omega)},
\end{equation}
where $S_n$ is the power spectral density of the noise on the induction coil.
$S_H$ is the power spectral density of the noise generated during the signal shaping and detection, while $\prod \mathcal H_i^{\ast} (\omega)$ represents the product of the conjugates of all response functions present in the readout process.
In the ADC-readout scenario, $\prod \mathcal H_i^{\ast} (\omega)$ corresponds to the conjugate of the circuit response function $\mathcal H^{\ast}$, and the $S_H$ is mainly influenced by the noise from the OPA.
In the magnetometer-readout scenario, $\prod \mathcal H_i^{\ast} (\omega)$ represents the combined conjugate response of both the circuit and magnetometer $\mathcal H^{\ast} \mathcal H^{\ast}_m$, and $S_H$ accounts mainly for the thermal noise from the Helmholtz coil.
The noise from the magnetometer is negligible.
Thermal noise originating from the induction coil significantly influences the overall noise characteristics, especially in the magnetometer-readout scenario.
This noise is modeled as Johnson-Nyquist noise~\cite{Johnson1928} \footnote{Note the expression of $S_n$ follows the convension in the field of signal processing~\cite{somalwar1988transient}.}:
\begin{equation}
    S_n (\omega) = 4k_B T R_{AC} (\omega) ,
\end{equation}
where $k_B$ is the Boltzmann constant, and $T$ is the temperature.
It is essential to emphasize that the thermal noise in this particular scenario does not exhibit the characteristic of ``white'' noise, which is typically assumed to have a frequency-independent power spectral density.
Due to the presence of a non-trivial alternating resistance in the induction coil, the thermal noise power increases with higher frequencies.
Fig.~\ref{fig:before_after_of} shows some waveform examples before and after applying OF for both the ADC-readout and magnetometer-readout scenarios.
In order to reduce the spectral waveform distortions due to the limited-length time window, specific-shaped time windows, such as the Hamming window~\cite{oppenheim1999discrete}, are introduced in the signal and noise processing.

The typical SNR is calculated under the assumption that the MM passes perpendicularly through the coil center with the speed of 10$^{-5}$ light speed (denoted as SNR$_0$ in the text).
After optimizing the circuit configuration for each test coil, the SNR$_0$s of each coil can be found in Table~\ref{tab:coils}.
It is worth noting that the SNR  depends on the MM's speed.
Fig.~\ref{fig:snr_dependence_on_speed} displays the SNR's velocity dependence of the prototype induction coils in both ADC-readout and magnetometer-readout modes.
The SNR increases with the increase in MM speed. 
However, SNR gradually tends toward saturation because the alternating resistance increasingly becomes significant at high frequency and the resonant frequency is preliminarily set to about several kHz in this study.
Increasing the resonant frequencies can potentially improve the SNR at the high velocity ranges.

\section{Validation of the signal simulation}
\label{sec:validation}

A validation test is performed to assess the accuracy and reliability of the signal simulation framework. 
The test mainly aims to validate the waveform amplitudes and shapes of the MM signal and the  thermal noise. 
These characteristics of the signal waveforms basically define the SNR$_0$ and  are crucial in the prediction of the sensitivity to the single Dirac MM.

\subsection{Signal validation}

The MM signal validation involves the utilization of a long-thin stimulation coil to generate a pulsed magnetic field that emulates an induction signal that could be caused by large numbers of Dirac MMs on the test coils.
The highest SNR$_0$ among the three test coils with ADC readout is about 0.57, mainly limited by the thermal noise of the induction coil and the noise of OPA, as shown in Table~\ref{tab:coils}.
On the contrary, the highest SNR$_0$ with magnetometer readout can reach 1.92 because of the low noise level of the magnetometer.
The stimulation coil utilized in the validation has a length of 50\,cm and a diameter of 10\,mm. 
The turns number density amounts to approximately 100 per centimeter. 
During the testing, the stimulation coil passes through the center of the induction coil, perpendicular to its coil plane.


In our experimental setup, we generate a voltage pulse with a square wave shape using a pulse generator.
Due to the prevalence of the electromagnetic noise in the surrounding environment and the limited precision of the pulse generator, it is not practically feasible to accurately emulate and test the signal response to a single Dirac MM. 
The generated magnetic flux is approximately 1.7$\times$10$^{-8}$\,Wb, equivalent to the flux created by about 4M Dirac MMs.
A resistor with a resistance of about 20\,$\Omega$ is connected in series with the stimulation coil.
The voltage drop across this resistor is monitored using a digitizer with a sampling rate of 2\,MHz, which is connected in parallel to the resistor. 
This allows us to model the microcurrent passing through the stimulation coil. 
It should be noted that in our experimental setup, we assume there are no leak fields associated with the tightly wound stimulation coil. 

\begin{figure*}[htp]
    \centering
    \includegraphics[width=0.9\textwidth]{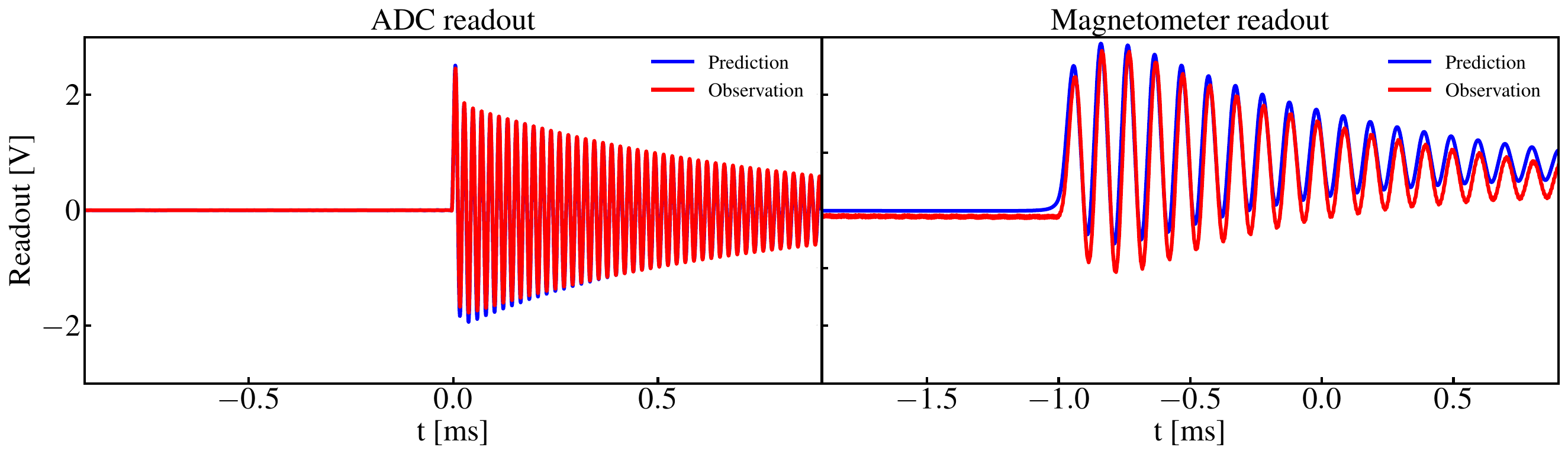}
    \caption{
    The measured and predicted signal waveforms are shown in red and blue solid lines, respectively.
    The left panel shows the results for ADC readout scenario, and the right panel shows for the magnetometer readout scenario.
    }
    \label{fig:signal_test_results}
\end{figure*}

Such test is performed for all three test induction coils with the ADC-readout scenario.
For the magnetometer-readout scenario, the V2 coil is tested which is expected to have the largest SNR among all three  coils.
Fig.~\ref{fig:signal_test_results} shows the comparison between the measured and predicted test signals in the time domain.
The readout signals can be parameterized as:
\begin{equation}
    S(t) = A \sin(\omega t + \phi) \cdot e^{-t/\delta},
\end{equation}  
where the $A$, $\omega$, $\delta$, and $\phi$ are the amplitude, frequency, decay rate, and phase, respectively.
The $A$, $\omega$, and $\delta$ are compared between the expectation and the measurement.
The results of the comparison are summarized in Table~\ref{tab:comparison}.
The measured frequencies and decay rates are consistent with the predictions, with bias no more than 0.3\% and 8.7\%, respectively.
This validates our response function models of the circuit and magnetometer.
The largest amplitude differences observed between measurements and predictions are about 12.5\% for the ADC readout and 9.2\% for the magnetometer readout.
This is considered to be likely due to the leak fields and uneven turn density of the stimulation coil.
Particularly the field leakage is more severe for V3 coil since the size of V3 coil is the largest among the tested ones.
In addition, the lower amplitude seen in the measurement with the magnetometer readout could be also due to the potential bias of the effective Lorentzian response shown in Eq.~\ref{eq:magnetometer_response}.
\textcolor{purple}{
}

\begin{table}[htp]
    \centering
    \begin{tabular}{c|c|c|c|c}
    \hline\hline
    \multirow{2}{*}{Coil}                               &
    \multicolumn{3}{c|}{ADC readout}                    &
    \multicolumn{1}{c}{Mag. readout}                    \\
    \cline{2-5}
     & V1 & V2 & V3 & V2                     \\
    \hline\hline
    A$_{\textrm{msr}}$ [V]                                  &
    1.830                                                  &
    1.906                                                   &
    1.143                                                   &
    3.899                                                  \\
    A$_{\textrm{prd}}$ [V]                                 &
    1.880                                                   &
    1.831                                                   &
    1.306                                                   &
    4.293                                                   \\
    A$_{\textrm{msr}}$/A$_{\textrm{prd}}$               &
    0.973                                                 &
    1.041                                                 &
    0.875                                                 &
    0.908                                                 \\
    $\omega_{\textrm{msr}}$ [kHz]                             &
    $58.8$                                                 &
    $296.5$                                                 &
    2.0                                                &
    $61.4 $                                             \\
    $\omega_{\textrm{prd}}$ [kHz]                            &
    $58.9$                                                 &
    $297.1$                                                 &
    2.0                                                 &
    $61.4$                                            \\
    $\omega_{\textrm{msr}}$/$\omega_{\textrm{prd}}$               &
    0.999                                                 &
    0.998                                                 &
    1.003                                                 &
    0.999                                                 \\
    $\delta_{\textrm{msr}}$ [ms]                             &
    $1.058$                                                 &
    $0.703$                                                 &
    10.309                                                 &
    $0.863$                                             \\
    $\delta_{\textrm{prd}}$ [ms]                           &
    $1.077$                                                 &
    $0.770$                                                 &
    10.886                                                 &
    $0.874$                                               \\
    $\delta_{\textrm{msr}}$/$\delta_{\textrm{prd}}$               &
    0.983                                                 &
    0.913                                                 &
    0.947                                                 &
    0.987                                                 \\
    \hline\hline
    \end{tabular}
    \caption{
    The ratios of the measured parameters versus the predicted ones.
    The parameters include the amplitude $A$, the resonant frequency $\omega$, and the decay rate $\delta$.
    The comparisons are performed for all three test induction coils (V1, V2, and V3) with the ADC readout scenario.
    The results of V2 test with the magnetometer readout scenario are shown.
    }
    \label{tab:comparison}
\end{table}

\subsection{Noise validation}

\begin{figure*}[htp]
    \centering
    \includegraphics[width=0.85\textwidth]{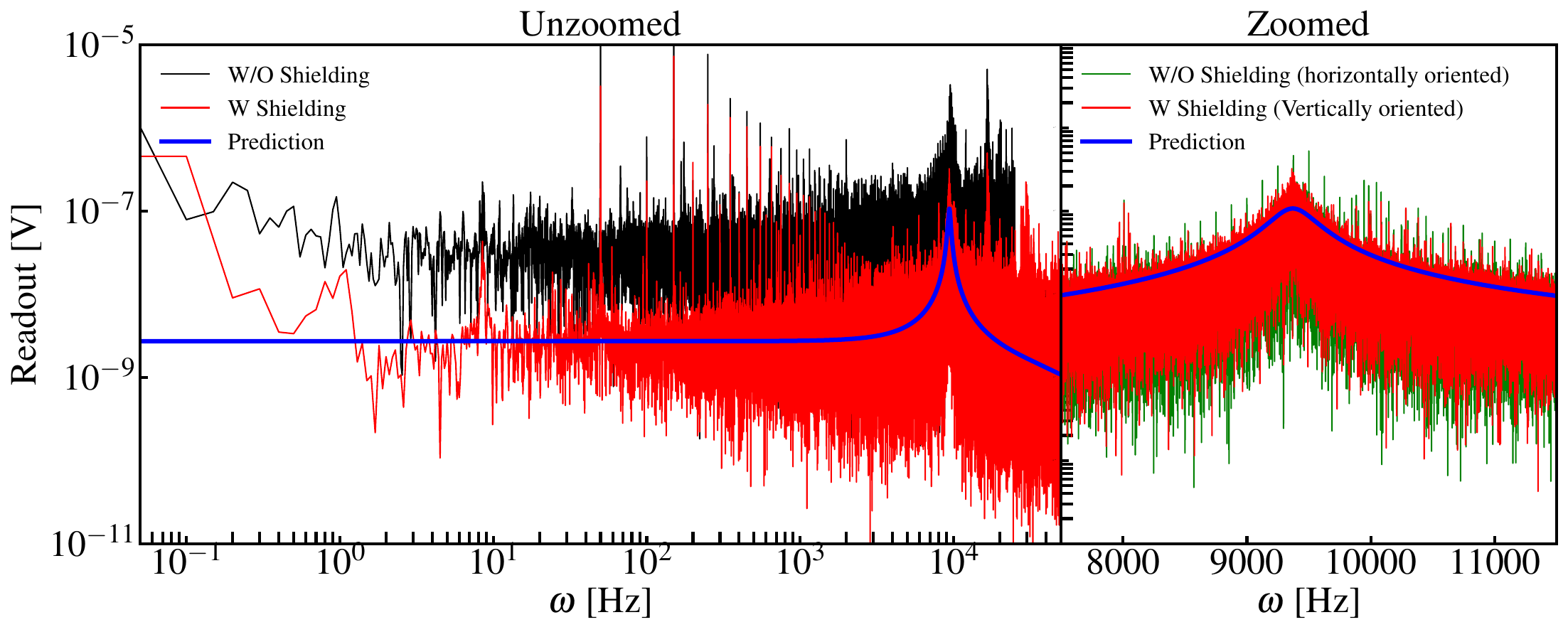}
    \caption{
    The noise spectra with different setups.
    The left panel shows the unzoomed spectra with and without copper box as the electromagnetic shielding, in black and red solid lines, respectively.
    The blue solid line gives the predicted thermal noise spectrum.
    The right panel gives a zoomed view of the spectra with shielding, around the resonant frequency of the induction coil. 
    The green solid line represents the noise spectrum when the induction coil is arranged so that its axis is oriented horizontally.
    }
    \label{fig:noise_validation}
\end{figure*}

To determine the intrinsic thermal noise power spectrum,  V1 coil is enclosed within a grounded metal box constructed of copper, which served as a Faraday cage.
Fig.~\ref{fig:noise_validation} displays the power spectra of  V1 coil under two conditions: when the coil is exposed to the air and when it is sealed inside the grounded copper box.
A significant reduction in noise is observed when the coil is enclosed in the copper box, indicating the presence of a strong electromagnetic noise background in the air.
Furthermore, the frequency domain analysis revealed distinct peak-like structures upon placing the coil inside the copper box.
These peaks corresponded to the multiples of the common frequency in utility, suggesting the presence of the leaked-in electromagnetic waves within the copper box, likely originating from the signal connectors.
This hypothesis is supported by the observation that the orientation of the induction coil influences the level of noise detected.
The lowest noise level is observed when the coil axis is in a vertical position, as shown in Fig.~\ref{fig:noise_validation}.
The observed noise frequency spectrum closely resembled the predicted one by the simulation, with a slightly lower amplitude (8.0\%) at the resonant frequency.

\section{Sensitivity to Cosmic Magnetic Monopole}
\label{sec:sensi_to_mm}

The search for the MMs will be conducted using an array of induction coils. 
The top and bottom of the coil array will be equipped with PSs, as depicted in Fig.~\ref{fig:scep_schematic_diagram}.
We take the detector situated at the sea level on Earth as the benchmark.
The dominant background is the pileup between the scintillation signals caused by the cosmic ray induced background particles (mainly muons and protons) in the PSs and the thermal noise in the induction coils.
The impact of this background can be mitigated by requiring more layers of the induction coils and the particle detectors.

To assess the sensitivity of the detector array to the cosmic MMs, we employ a simple ideal configuration. 
This configuration consists of induction coils with a diameter of 12\,cm (same as V2 coil), arranged vertically and compactly instrumented. 
The array's size is assumed to be sufficiently large to disregard any edge effects.
The alternating resistance's frequency dependence of each induction coil is assumed to followed the one of V2 coil, and each coil is assumed to have negligible height.
In this analysis, a simple over-threshold trigger is conducted on waveform of each induction coil after the OF applied.
The coil array is equipped with the PS layers at the top and bottom, and these layers are positioned approximately 1 meter apart.
Each PS layer is composed of two sets of PS panels arranged perpendicular to each other. 
This arrangement allows for the reconstruction of events' transverse positions.

\subsection{Acceptance to cosmic MM induction}

The cosmic MM is assumed to exhibit isotropic behavior.
However, due to the geometry of the induction coil, there is an inherent acceptance loss of (1-$\pi$/4) for each layer of coils.
We consider simply the coil layers are identical and sufficiently close to each other, so that we can consider such setup having a conservative acceptance loss of (1-$\pi$/4) due to coil geometry.
Optimizing the coil geometry and arrangement between layers can alleviate the acceptance loss to some extent.

\begin{figure}
    \centering
    \includegraphics[width=0.90\columnwidth]{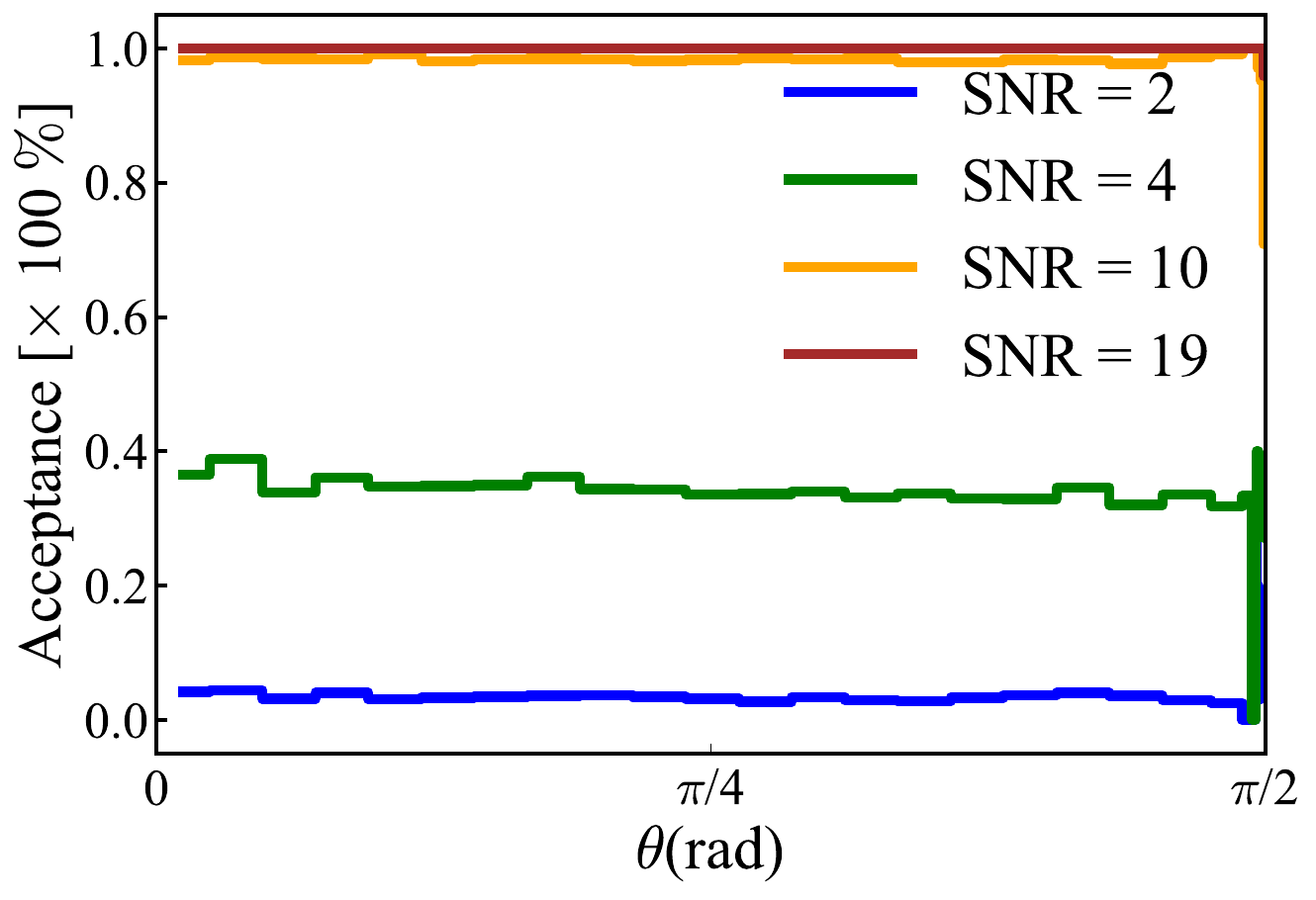}
    \caption{
    The acceptance to a single Dirac MM $A_{\rm n}$ as a function of transverse angle($\theta$) at a fix threshold $\alpha$ that makes the mis-triggering noise rate to be 6.8Hz.
    The blue, green, yellow, and red solid lines give the MM acceptances with assumed SNR$_0$ of 2, 4, 10, and 19.} 
    \label{fig:acceptance_theta}
\end{figure}

The dependence of the SNR on the transverse distance of the MM to the coil center is weak.
In Fig.~\ref{fig:acceptance_theta}, the average acceptance to cosmic MM is displayed as a function of the polar angle ($\theta$), considering various assumptions regarding the SNR$_0$.
Only when the $\theta$ approaches $\pi/2$, the acceptance drops quickly.

\begin{figure}
    \centering
    \includegraphics[width=0.90\columnwidth]{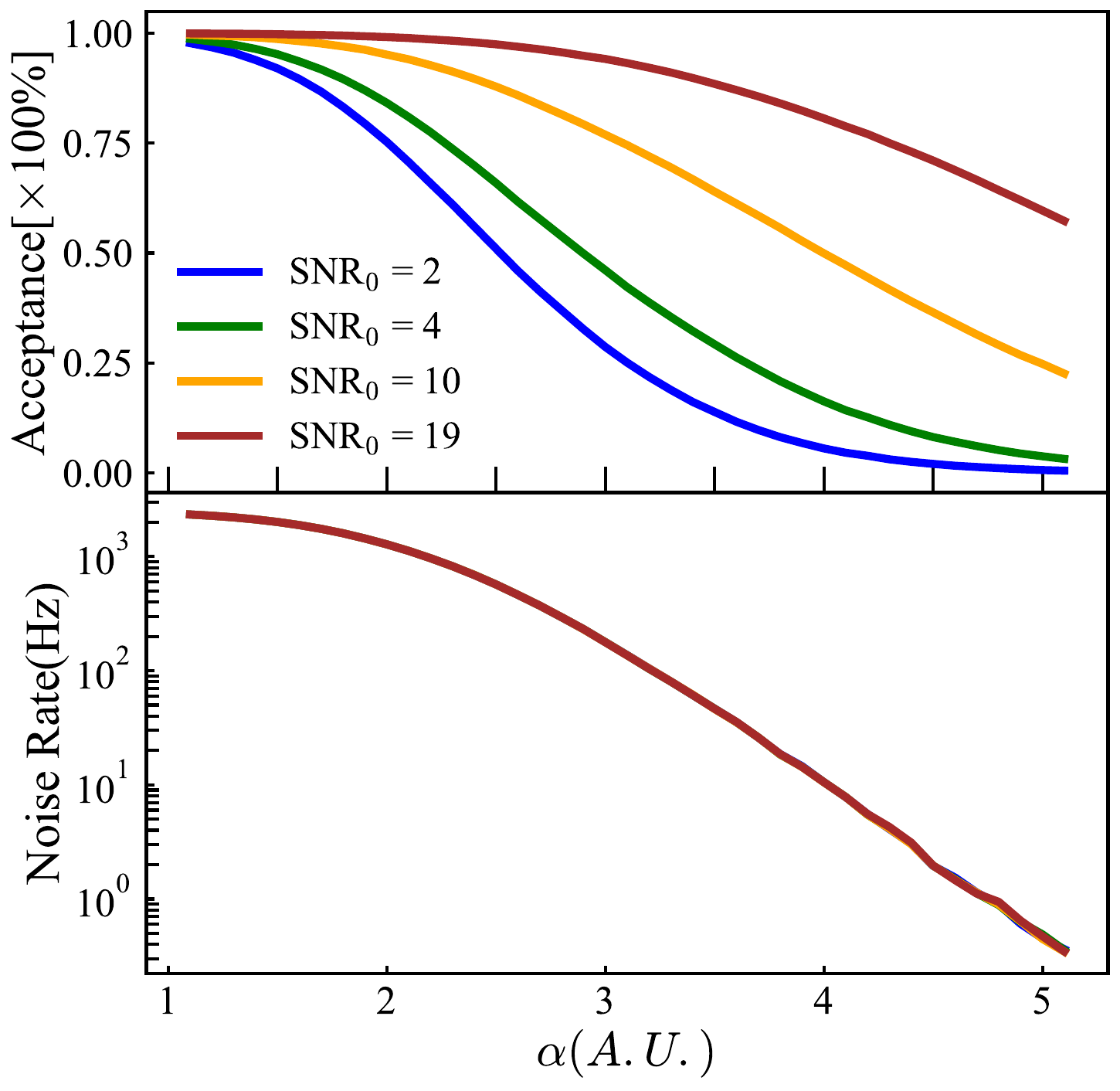}
    \caption{
    The acceptance to a single Dirac MM $A_{\rm n}$ and the mis-triggered thermal noise rate $R_{\rm n}$  on a single coil as a function of the trigger threshold $\alpha$ are shown in the top and bottom panels, respectively.
    The blue, green, yellow, and red solid lines in the top panels give the MM acceptances with assumed SNR$_0$ of 2, 4, 10, and 19.} 
    \label{fig:mm_acceptance}
\end{figure}

\subsection{Background of Induction Signal}
\label{subsec:induction_background}

The cosmic rays and their secondaries, such as high-energy protons, muons, and electrons, that are common in the terrestrial  environment, deposit energy in the top and bottom PSs, but produce negligible induction signals.
These particles possess magnetic dipoles and travel mostly at relativistic speeds, resulting in a distinct resonant induction pulse shape and faster time response compared to those from the MMs.
Therefore, we consider that relativistic cosmic rays and their secondaries do not produce any significant background for the induction signals.

However, the energy depositions detected by the PSs may coincide with the abundant thermal noise present in the induction coils. 
As discussed in Subsection \ref{subsec:recon_and_thermal_noise}, the thermal noise arises from the non-zero alternating resistances of the induction coil and constitutes the main background for the MM induction signal search.
In this analysis, a simple over-threshold approach is employed as the trigger method on a single coil. 
The dependence of the acceptance and noise rate on the threshold (denoted as $\alpha$ in the text) with different assumed $\textrm{SNR}_0$s are shown in Fig.~\ref{fig:mm_acceptance}.
The rate of the mis-triggering of thermal noise $R_{\textrm{ind}}$ and the acceptance across the coils to form a track-like event $A_{\textrm{ind}}$ can be expressed as follows:
\begin{equation}
\left\{
\begin{aligned}
    R_{\rm ind}(\alpha) & =  \frac{(R_{\rm n}(\alpha) \Delta t)^{N_{\rm c}-1}}{N_{\rm c} !}\cdot R_{\rm n}(\alpha) \\
A_{\rm ind}(\alpha) & =  \left( A_{\rm n}(\alpha)\right)^{N_{\rm c}}
\end{aligned}
\label{eq:coils'acceptance}
\right.
\end{equation}
Here, $R_{\rm n}(\alpha)$ and $A_{\rm n}({\alpha})$ represent the mis-triggered noise rate and acceptance of a single induction coil at a given threshold $\alpha$. 
$N_c$ denotes the number of coils required to detect the induction trigger (coincidence number). 
$\Delta t$ represents the time response of the induction signal, which is related to the resonant frequency of the induction coil. For the analysis, we assume $\Delta t = 100\,\mu s$.

\subsection{Background of Scintillation Signal}
\label{subsec:ionization_background}

\begin{figure}[htp]
    \centering
\includegraphics[width=0.9\columnwidth]{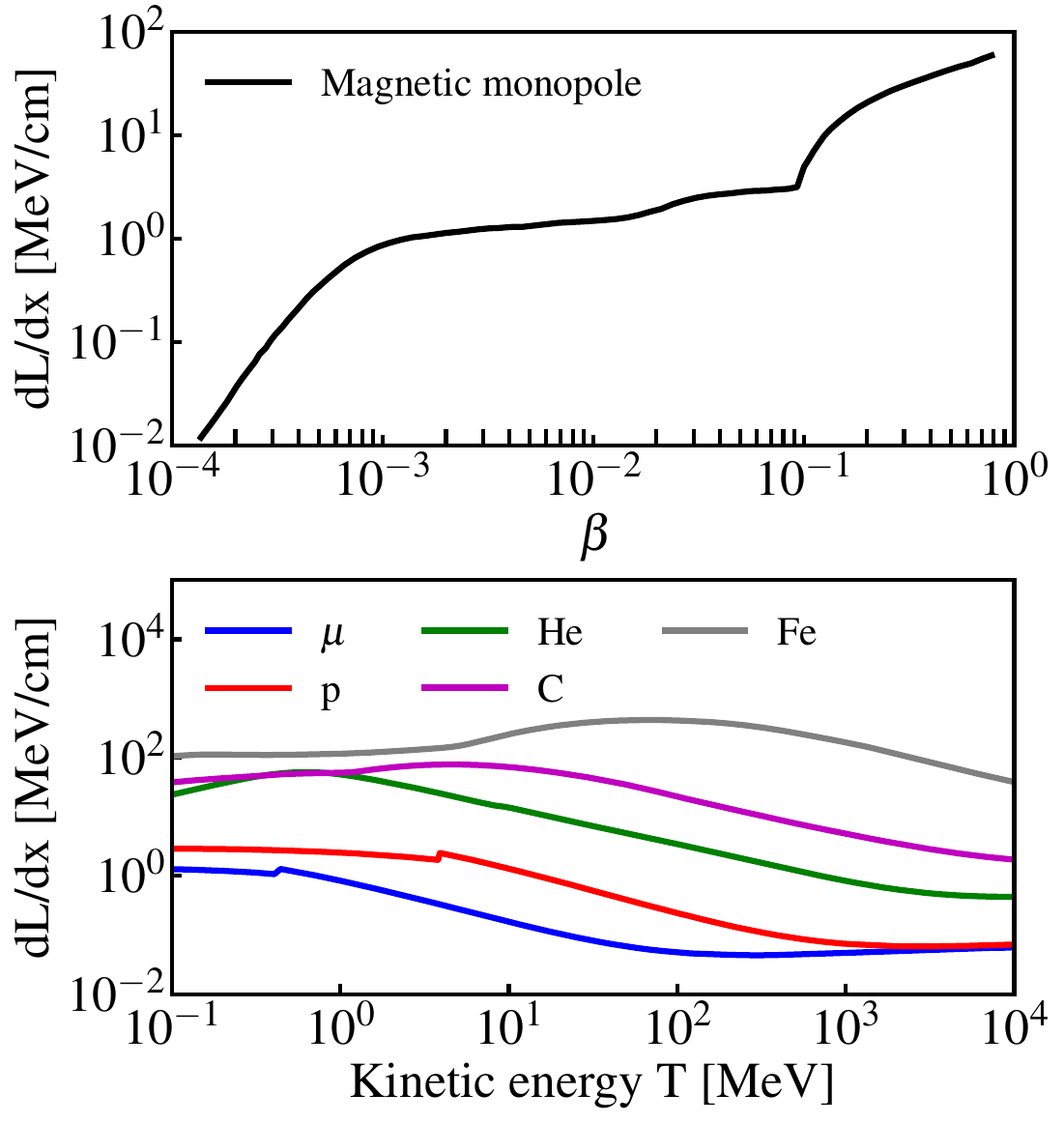}
    \caption{
    The light yield per distance ${\rm d}{\it  L}/{\rm d}{\it x}$ of a Dirac MM in the PS as a function of the MM speed, based on~\cite{derkaoui1999energy}, is shown in the upper panel.
    The d$L$/d$x$ of the muon, proton, helium, carbon, and iron nucleus are shown in the lower panel.
    The d$L$/d$x$ of muon is calculated based on the stopping power d$E$/d$x$ from PDG~\cite{groom2001muon}.
    The d$E$/d$x$ of proton, helium, carbon, and iron nucleus are from PSTAR and ASTAR database~\cite{international1993stopping}.
    }
    \label{fig:stopping_power}
\end{figure}

\begin{figure}[htp]
    \centering
    \includegraphics[width=0.9\columnwidth]{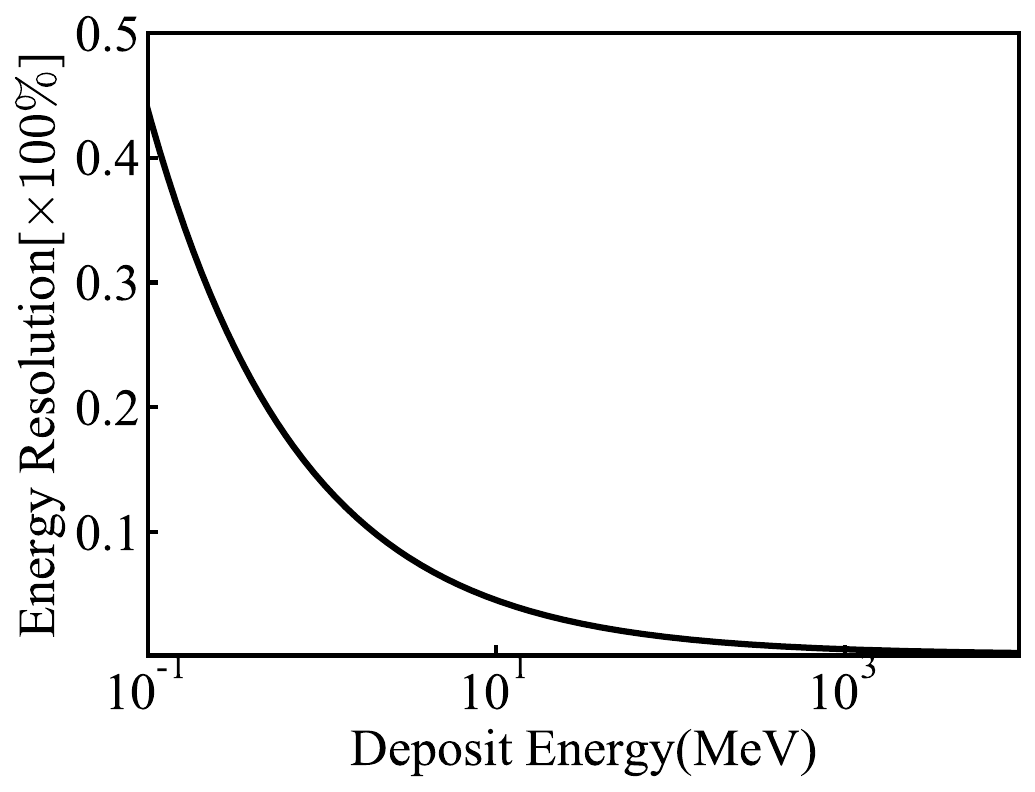}
    \caption{
    The energy resolution as a function of total deposit energy of the plastic scintillator.
    }
    \label{fig:energy_resolution_ps}
\end{figure}

\begin{figure}[htp]
    \centering
    \includegraphics[width=0.9\columnwidth]{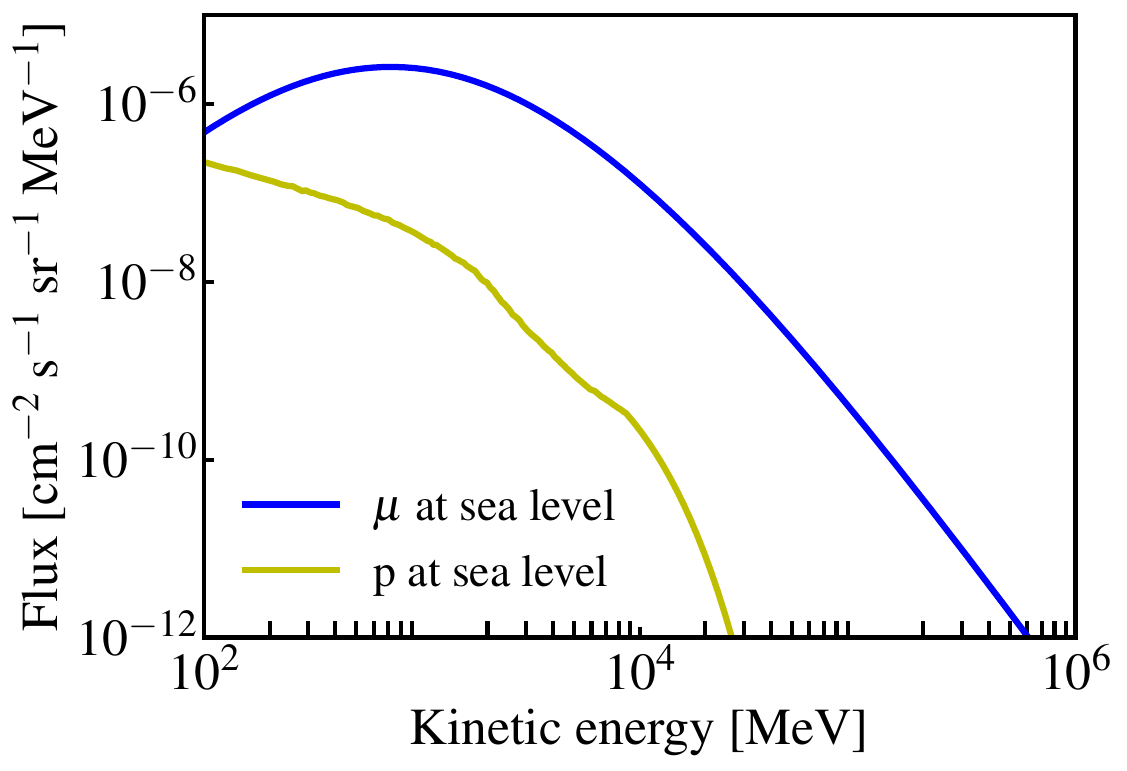}
    \caption{
    The fluxes of background particles as a function of their kinetic energies.
    The blue and yellow solid lines represent the muon and proton fluxes at sea level, which are calculated based on Bugaev/Reyna model~\cite{su2021comparison} and simulated by CRY algorithm~\cite{hagmann2007cosmic}, respectively.
    }
    \label{fig:particle_background_fluxes}
\end{figure}


The rate of the reconstructed scintillation signals on the PSs is mainly affected by two factors: random pileups occurring between the top and bottom PSs, and the passage of a relativistic particle through both PSs.
The energy threshold of PS is estimated to be $\sim$0.1\,MeV, below which the contribution of SiPM dark count pileup skyrockets.
This reconstructed scintillation signal necessitates the presence of two energy depositions, one on each of the top and bottom PSs.
It is crucial for the reconstructed energies, timings, and transverse positions of these two energy depositions to align with the expected energy, time of flight (ToF), and track characteristics of the MM of interest. 
The differential reconstructed scintillation signal rate per unit area per radian on the two PS panels can be expressed as the sum of two components: the rate arising from pileup events, denoted as $\mathcal{R}_{pile}$, and the rate resulting from direct passage of particles, denoted as $\mathcal{R}_{part}$.
\begin{equation}
\left\{
\begin{aligned}
\mathcal{R}_{\rm pileup} (\theta) & = \frac{1}{2} \left( \int  \mathcal{R}_{\rm ion}(\theta^\prime) {\rm sin}(\theta^\prime) {\rm d}\theta^\prime \right)^2 \Delta t_{\rm PS} \frac{4\pi^3 d^2}{{\rm cos}^3 \theta}, \\
\mathcal{R}_{\rm part} (\theta) & = \int \epsilon^2 (E_0) \mathcal{F} {\rm d}E_0, \\
\mathcal{R}_{\rm ion} (\theta) & = \int \epsilon (E_0) \mathcal{F} {\rm d}E_0, 
\end{aligned}
\right.
\end{equation}
where $\Delta t_{\rm PS}$ is the pileup time window determined by the PS time resolution, which is assumed to be 10\,ns~\cite{zhou2020time}, and $d$=1\,m is the distance between top and bottom PSs.
$\mathcal{R}_{ion} (\theta)$ is the effective scintillation rate on a single PS layer given a zenith angle.
$\mathcal{R}_{pile}$, $\mathcal{R}_{part}$, and $\mathcal{R}_{ion} (\theta)$ are in unit of cm$^{-2}$s$^{-1}$sr$^{-1}$.
We require $\mathcal{R}_{\rm ion}$ to be the rate after an energy range cut that has a coverage of 99.5\% (3$\sigma$) for the MMs.
Such cuts give an effective efficiency to background particles of $\epsilon (E_0)$, which also applies to direct particle passage but with two layers of PSs both requiring such energy selection.
The $\mathcal{F}$ is the particle flux.
Note that $\epsilon (E_0)$ depends on the speed of MM. 
The total scintillation background rate per unit area that takes all the angles into account can be expressed as: 
\begin{equation}
    R_{\rm ps}(\beta) =  2 \pi \int_0^{\theta_{\rm \textrm{max}}} (\mathcal{R(\theta)}_{\rm pileup} + \mathcal{R(\theta)}_{\rm part}) {\rm sin}(\theta) {\rm d}\theta.
\end{equation}
In order to avoid the numerical infinity of $\mathcal{R}_{\rm pileup}$ when $\theta$ is $\pi/2$, we set the $\theta_{\textrm{max}} = 80^{\circ}$.

In practice, the value of $\mathcal{R}_{ion}$ is influenced by several factors, including the background particle flux and spectrum, the ability to determine the direction of the MM using the induction signals, and the energy resolution of the PS.
Also, in order to determine the $\epsilon(E_0)$, the amount of light produced in the PS by a Dirac MM is needed, which depends on the MM velocity. 
The detectable stopping power, also known as the light yield, of a Dirac MM on the PS as a function of MM speed is presented in the top panel of Fig.~\ref{fig:stopping_power}, based on the calculations in Ref.~\cite{derkaoui1999energy}.
To differentiate the energy deposition of the MM from common background particles such as protons, electrons, alpha particles, and muons, we require that the reconstructed energy falls within 3 times the energy resolution.
The intrinsic energy resolution of the PS, as a function of the total deposited energy, is obtained through an optical simulation based on GEANT4~\cite{agostinelli2003geant4}. 
The energy resolution is illustrated in Fig.\ref{fig:energy_resolution_ps}.
The reconstruction resolution of the transverse position in the PS-based array primarily depends on the width of the PS panel, which is significantly smaller than the size of the induction coil. 
Consequently, the track reconstructed by the PS exhibits much higher resolution compared to the one reconstructed by the induction coils.
For this analysis, we conservatively consider $\mathcal{R}_{pile}$ after the coincidence requirement to be the background rate within a 12 cm-diameter circle, which corresponds to the size of the V2 coil used in the estimation.

In a terrestrial detector situated at sea level, the primary background particles are the atmospheric muons, as well as the residual high-energy protons.
Muons with kinetic energies ranging from hundreds of MeV to hundreds of GeV exhibit minimal ionizing behavior when interacting with matter, enabling them to easily traverse the surrounding materials near the detector, including the top and bottom PSs.
On the other hand, the proton flux experiences a significant reduction as it traverses the atmosphere due to the ionization and radiative processes. 
The protons leave a higher ionization density in the PS compared to the muons, approaching the ionization density that could be produced by the Dirac MMs within a certain range of speeds.
The stopping power $dE/dx$ and the light yields $dL/dx$ of protons and muons, corresponding to detectable energy ranges, in the PS are calculated based on the PSTAR and ASTAR databases~\cite{international1993stopping}, PDG sources~\cite{groom2001muon}, and the methodology outlined in Ref.~\cite{derkaoui1999energy}. 
These light yields are presented in Fig.\ref{fig:stopping_power}.
To model the flux and angular distribution of atmospheric muons at sea level, the Bugaev/Reyna model~\cite{su2021comparison} is employed, while the flux of high-energy protons is simulated using CRY algorithms~\cite{hagmann2007cosmic}. The fluxes are shown in Fig.\ref{fig:particle_background_fluxes}.

\begin{figure}[htp]
    \centering
    \includegraphics[width=0.9\columnwidth]{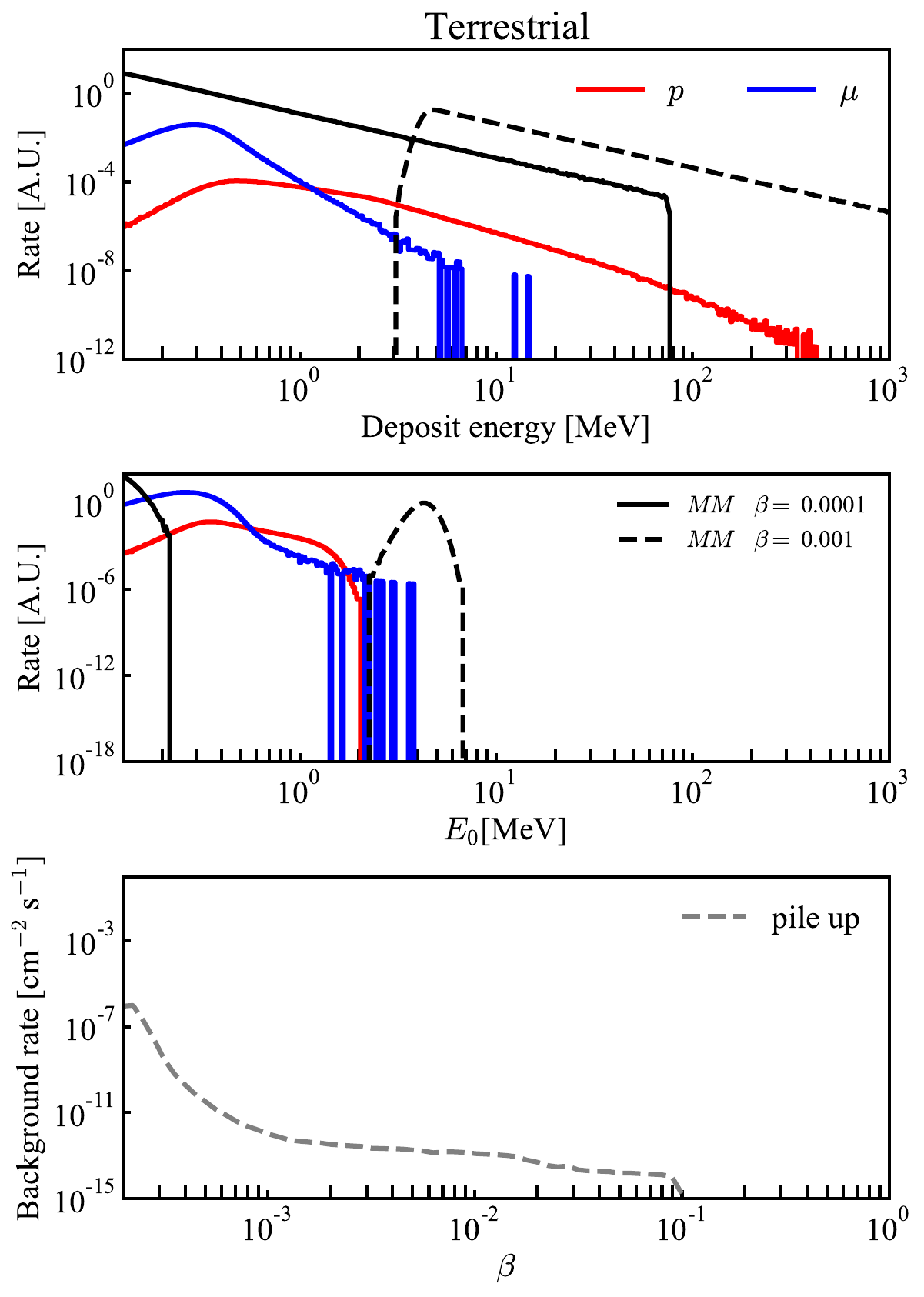}
    \caption{
    The simulated reconstructed energy distributions of the background particles and MMs with different speeds from all the angles are shown in the top panel.
    The middle panels show the $E_0$ distribution of background particles and MMs.
    The colored solid lines showing the contributions from different background particles, while the black solid and dashed lines show the distributions of MMs with $\beta$ of 0.0001 and 0.001, respectively.
    The lower panel shows the background rate after the energy and ToF requirements, including the direct passage component (colored dashed line) and pile up (grey line) component, as a function of assumed MM speed. }
    \label{fig:background_rate_vs_speed}
\end{figure}


We performed a toy Monte Carlo (MC) simulation to calculate the background and MM energy deposition spectra. 
The simulation takes into account the spread in the deposited energy caused by the variation in particle traveling lengths within the PS due to different incoming particle angles.
In the final analysis of the top and bottom PSs, we reconstruct the zenith angles of the incoming background particles, MMs, or "fake" particles reconstructed from pileup events.
All the deposited energies are corrected to the equivalent energy deposition when the particle passed through the material perpendicularly, denoted as $E_0$.
The top panel of Fig. \ref{fig:background_rate_vs_speed} illustrates the predicted deposited energy spectra of Dirac MMs with different assumed speeds and the background particles in a single PS layer (before $E_0$ correction). The middle panel gives the $E_0$ differential rates for a single PS layer. The lower panel depicts the total scintillation background rate $R_{ps}$ for different assumed MM speeds, requiring that we select an $E_0$ range that covers 99.5\% (3$\sigma$) of the MMs and a ToF that also covers 99.5\% (3$\sigma$) of the MMs.
Considering the energy threshold of 0.1\,MeV in the PS, the acceptance to MMs with speeds lower than about 2.5$\times$10$^{-4}$ light speed is negligible.


\subsection{Total Background and Sensitivity}
\label{subsec:sensitivity}

\begin{figure*}[ht]
    \centering
    \includegraphics[width=0.95\textwidth]{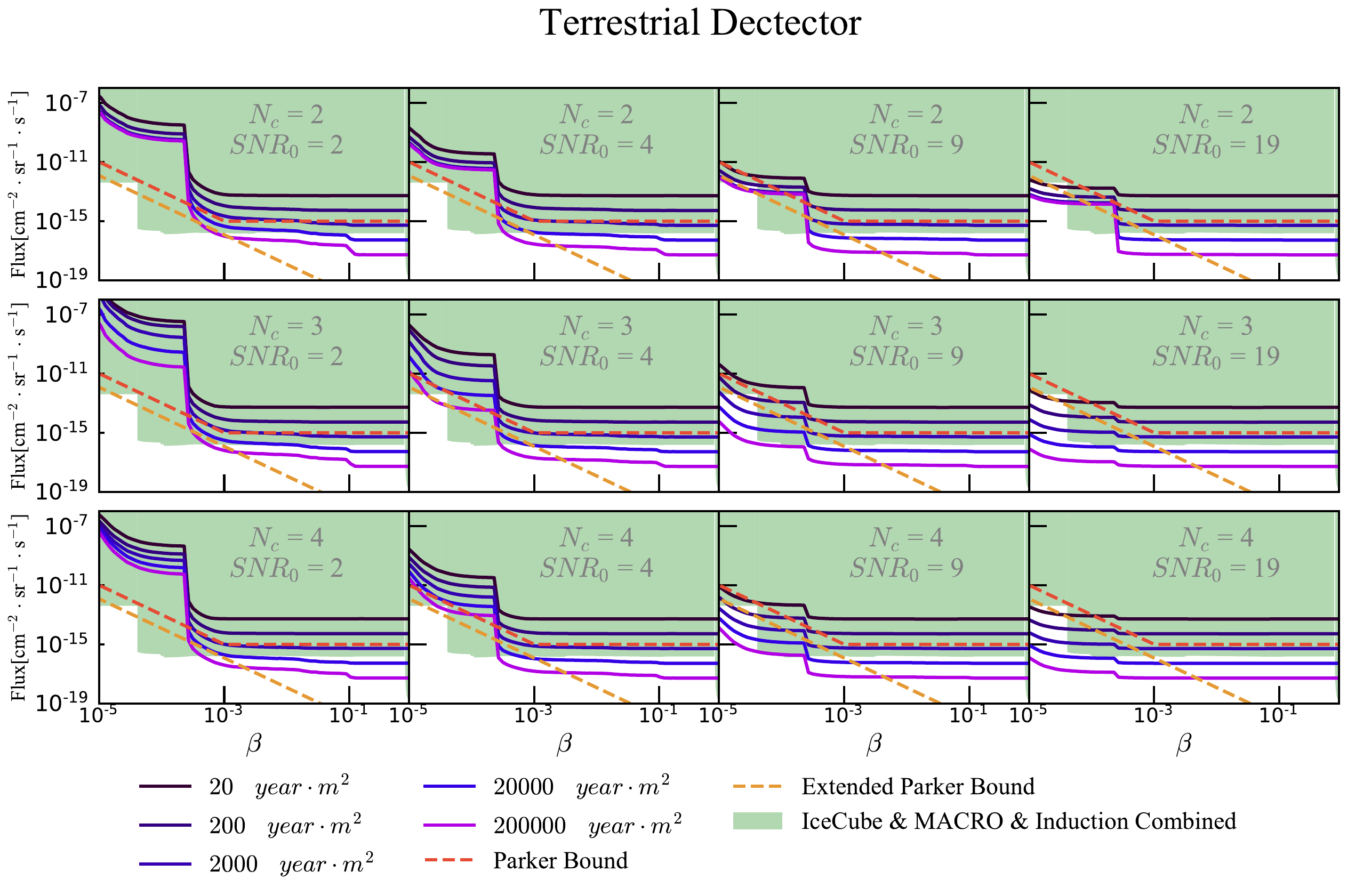}
    \caption{
    The projected 90\% exclusion sensitivity $\mathcal{Q}$ of the cosmic MM flux for a terrestrial detector. 
    Each sub-figure in the table shows the projected sensitivity as a function of the MM speed.
    The colored solid lines represent the sensitivities under different exposure as illustrated in the legends.
    The green region shows the flux constraints that have been given by MARCO~\cite{macro2002final}, IceCube~\cite{iceCube2022_mm_result} and superconductivity induction experiments~\cite{giacomelli1984magnetic}. 
    Each row shows the sensitivity with different SNR$_0$ assumption and each column gives the sensitivities with different $N_{\rm c}$.
    The mass of the MM is assumed to be sufficiently high so that the acceptance loss due to the earth shielding is negligible.
    }
    \label{fig:earth detector}
\end{figure*}

Low-speed MMs ($\beta<2.5\times10^{-4}$) are unable to produce enough lights to surpass the energy threshold in PS.
Searches of such MMs need to be performed with induction signal only. 
The total background rate per unit area can be expressed as:
\begin{equation}
R(\alpha,\beta) = 
\left\{
\begin{aligned}
    & R_{\rm ind} (\alpha) \cdot R_{\rm ps}(\beta) \cdot \Delta t ; \quad \beta >2.5\times 10^{-4} \\
    & \frac{ R_{\rm ind}(\alpha)}{\pi r_{\rm coil}^2} ; \quad \beta <2.5\times 10^{-4}.
\end{aligned}
\right.
\label{eq:final rate}
\end{equation}
Here $r_{\rm coil}$ denotes as the radius of a single coil, and we set it to be 6 cm in the following calculation.
The $R_{\rm ind}(\alpha)$ and ${R}_{\rm ps}(\beta)$ are given in Subsec.~\ref{subsec:induction_background} and~\ref{subsec:ionization_background}, respectively.
The induction threshold $\alpha$ is optimized based on the sensitivity, which is equivalently the mean Feldman-Cousins upper limit~\cite{feldman1998unified}  $\mathcal{Q}$  under the background-only hypothesis.
The optimized threshold depends on the SNR$_0$, the total exposure, the assumed MM speed $\beta$ and the coincidence number:
\begin{equation}
    \mathcal{Q} =\frac{\sum_i \mathcal{P}(i,\Lambda R(\alpha, \beta) \cdot \mathcal{F}(i, \Lambda R(\alpha,\beta))}{4\pi A_{\rm ps}(\beta) A_{\rm ind}(\alpha) A_{\rm \textrm{geo}} \Lambda},
\end{equation}
where $A_{\rm ps}(\beta)$ is the acceptance of PSs due to energy threshold and the angular cut-off $\theta_{\textrm{max}}$.
$A_{\rm ind}(\alpha)$ is the trigger acceptance of the induction coils from Eq.~\ref{eq:coils'acceptance}, and $A_{\textrm{geo}}=\pi/4$ is the geometric acceptance of the induction coils to the MMs.
$\Lambda$ is the assumed exposure.
$\mathcal{F}(x, y)$ denotes as the Feldman-Cousins upper limit  at $90\%$ CI under background only hypothesis, when $x$ events are observed with $y$ background predicted.
$\mathcal{P}(x, y)$ is the Poisson probability of observing $x$ events under predicted background of $y$.

The estimated 90\% exclusion sensitivities of the MM flux as a function of the MM speed for a terrestrial detector are given in Fig.~\ref{fig:earth detector}, with different assumptions of the SNR$_0$, N$_c$, and exposure. 
And we also give the sensitivities for a moon-based detector in the appendix. 
The most stringent constraints to the MM flux at different speed ranges from all the induction experiments~\cite{giacomelli1984magnetic}, MACRO~\cite{macro2002final}, and IceCube~\cite{iceCube2022_mm_result, abbasi2022search}, as well as the astrophysical constraints (Parker boudary~\cite{parker1970origin} and extended Parker boundary~\cite{adams1993extension, lewis2000protogalactic}), are also plotted for comparison.
SCEP has the potential to achieve excellent background suppression for cosmic Dirac MMs traveling at speeds exceeding $\sim$ 2.5$\times$10$^{-4}$ light speed.
The sensitivities within this speed range are primarily dominated by the exposure (the product of exposure time and area of detection area).
For MMs traveling at speeds below approximately 2.5$\times$10$^{-4}$ light speed, the Dirac MMs are unable to produce scintillation lights in the PSs, causing SCEP to operate solely in induction-only search mode.
As a result, sensitivities in this speed range decrease due to the absence of the scintillation/induction coincidence.
However, the use of a higher number of coil layers and a larger value of SNR can help recover the lost sensitivity to some extent. 
We estimate that a SNR$_0 > 4.5$ and a N$_c > 2$ are required to reach the current best limits by all the induction searches~\cite{giacomelli1984magnetic} with an exposure of 20000 year$\cdot$m$^2$.
Also, the sensitivity keeps decreasing as the speed of MM decreases due to the SNR's dependence on the MM speed.




\subsection{Acceptance Loss due to Earth Shielding}
\label{subsec:earth_shielding}

\begin{figure}
    \centering
    \includegraphics[width=0.95\columnwidth]{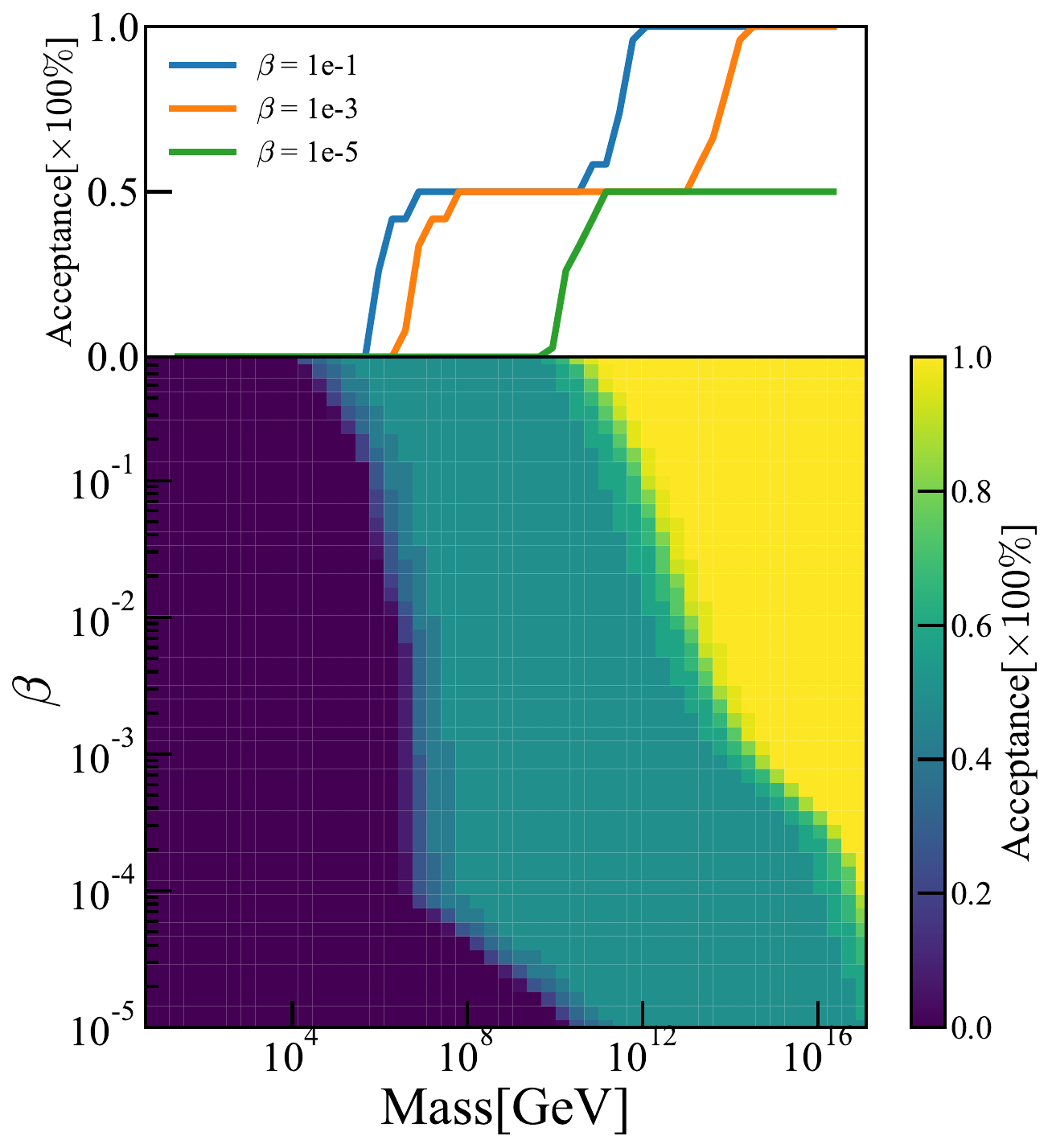}
    \caption{
    The acceptance of a sea-level terrestrial detector to the isotropic cosmic MMs due to the earth shielding effect is represented by the color in the bottom panel.
    The projected acceptances as a function of the MM masses at three MM speeds are shown in the top panels.
    }
    \label{fig:earth_shielding_acceptance}
\end{figure}

The Earth's matter can shield the cosmic MMs with insufficient mass, causing an additional acceptance loss for a terrestrial MM detector.
We perform a simple estimation of the acceptance loss due to this Earth shielding effect, assuming the Earth is a sphere composed of only the atmosphere, silicon mantle, and iron core layers.
The thicknesses of these three layers are taken as 2000, 3500, and 2900 km, respectively. 
The air density distribution along the altitude uses the data from Ref.~\cite{Characterisations}.
The energy losses of the MMs in the air, the silicon, and iron follow the model given in Ref~\cite{DERKAOUI1998173}. 
Particularly, the energy losses in the air for the MM $\beta$ in 10$^{-3}$ to 10$^{-2}$ and 10$^{-5}$ to 10$^{-4}$ light speed are scaled from that of the a proton and of the liquid helium, respectively, same as in Ref~\cite{DERKAOUI1998173}.
We also neglect the scattering of the MMs inside the Earth and the atmosphere.
The acceptance loss of the isotropic cosmic MMs due to the Earth shielding effect for a terrestrial sea-level detector is calculated and shown in Fig.~\ref{fig:earth_shielding_acceptance}.

For the upgoing cosmic MMs, they are not able to penetrate Earth's mantle and core if their masses are less than 10$^{11}$ to 10$^{15}$\,GeV (depending on the MM speed), leading to the reduction of the acceptance by about half.
When the MM mass is less than certain limit (for example, 10$^7$\,GeV for $\beta$ of 0.001), no cosmic MMs can be searched for using sea-level detectors since the atmosphere absorbs all of the downgoing MMs.

\section{Summary and Discussion}

The SCEP experiment aims to detect the induction signal and scintillation signal simultaneously when a cosmic MM passes through.
This dual-signal detection approach not only can provide smoking-gun signature for a potential cosmic MM traversal, but also suppresses the background caused by the environmental charged particles in the PSs and thermal noise in the induction coils.
This allows the experiment to be deployed in a room-temperature environment at sea-level (or even high-altitude) locations, which is more cost-effective compared to experiments requiring an underground environment or cryogenic detector conditions.
The SCEP design holds the potential to achieve a large-area detector array, which is crucial in reaching the MM flux sensitivity beyond the current experimental and astrophysical constraints.

We have developed a dedicated simulation framework for estimating the induction signal performance. This simulation has been validated through tests performed on three prototype induction coils with distinct geometric parameters, using both the magnetometer readout and the direct ADC readout.
The differences between the model predictions and measurements in terms of signal amplitudes, signal shapes, and noise amplitudes are within 10\%.
The thermal noise of the induction coil, which increases with frequency due to the alternating resistance, is the main contributor of the induction background. 
Based on the simulation of the induction signal and an estimation of the environmental particle backgrounds, we provide an assessment of the sensitivity to the cosmic MM flux. 
Thanks to the coincidence between the induction and scintillation signals, the SCEP experiment is able to achieve a ``background-free'' search mode for MM speeds greater than approximately 2.5$\times$10$^{-4}$ the light speed.
In this speed range, the background is significantly suppressed, and the sensitivity to the MMs is mostly proportional to the total exposure.
An exposure of about 20,000 year$\cdot$m$^2$ will be sufficient to reach the current flux limit set by the MACRO experiment~\cite{macro2002final} over a wide range of the MM speeds (from $\sim$2.5$\times$10$^{-4}$ to 0.75 light speed).
For MMs with speeds less than about 2.5$\times$10$^{-4}$ light speed, they are no longer able to create scintillation signals in the PSs.
The SCEP experiment will operate in an induction-only mode.
To reach the limits set by all previous induction-only experiments combined~\cite{giacomelli1984magnetic}, an SNR$_0$ greater than 4.5 and a number of coil layers more than 2 are preferred, given the assumed exposure of 20,000 year$\cdot$m$^2$.

The current three test coils, without optimization, have achieved an SRN$_0$$\sim$2.
Further optimization is required to reach the SNR$_0$ larger than 4.5.
To improve the SNR of the induction signal, two approaches are planned: 1) optimization of the induction circuit and 2) the use of materials with high magnetic permeability.
The latter is of particular potential.
A preliminary estimation shows that a high-permeability magnetic core can effectively amplify the MM signal by at least one order of magnitude, with the amplification factor depending on the core material and geometric dimensions. 
This amplification can be achieved without introducing significant additional thermal noise from the eddy currents and hysteresis within the frequency range of interest.
This would result in an overall amplification of the SNR by several hundred times.
Such a level of SNR is sufficient to meet the requirements for the planned large-area array detectors.
More details on the optimization and performance improvements will be provided in future dedicated publications. 
Similar analyses have also been performed in Ref.~\cite{tumanski2007induction,coillot2015magnetic}.

In addition to the induction signal optimization, increasing the number of the PS detector layers and incorporating different types of particle detectors hold promise for further reducing the background rate due to the environmental particles. 
Moreover, more advanced algorithms, such as those based on the deep neural networks, have the potential to enhance the background rejection by leveraging the full range of information obtained from the experimental data.

\section*{Acknowledgements}

This project is supported by grants from National Science
Foundation of China (No. 12250011), and by the Frontier Scientific Research Program of Deep Space Exploration Laboratory under grant No. 2022-QYKYJH-HXYF-013.

\appendix

\section{Sensitivity for a Moon-based Detector}

\begin{figure}[htp]
    \centering
    \includegraphics[width=0.95\columnwidth]{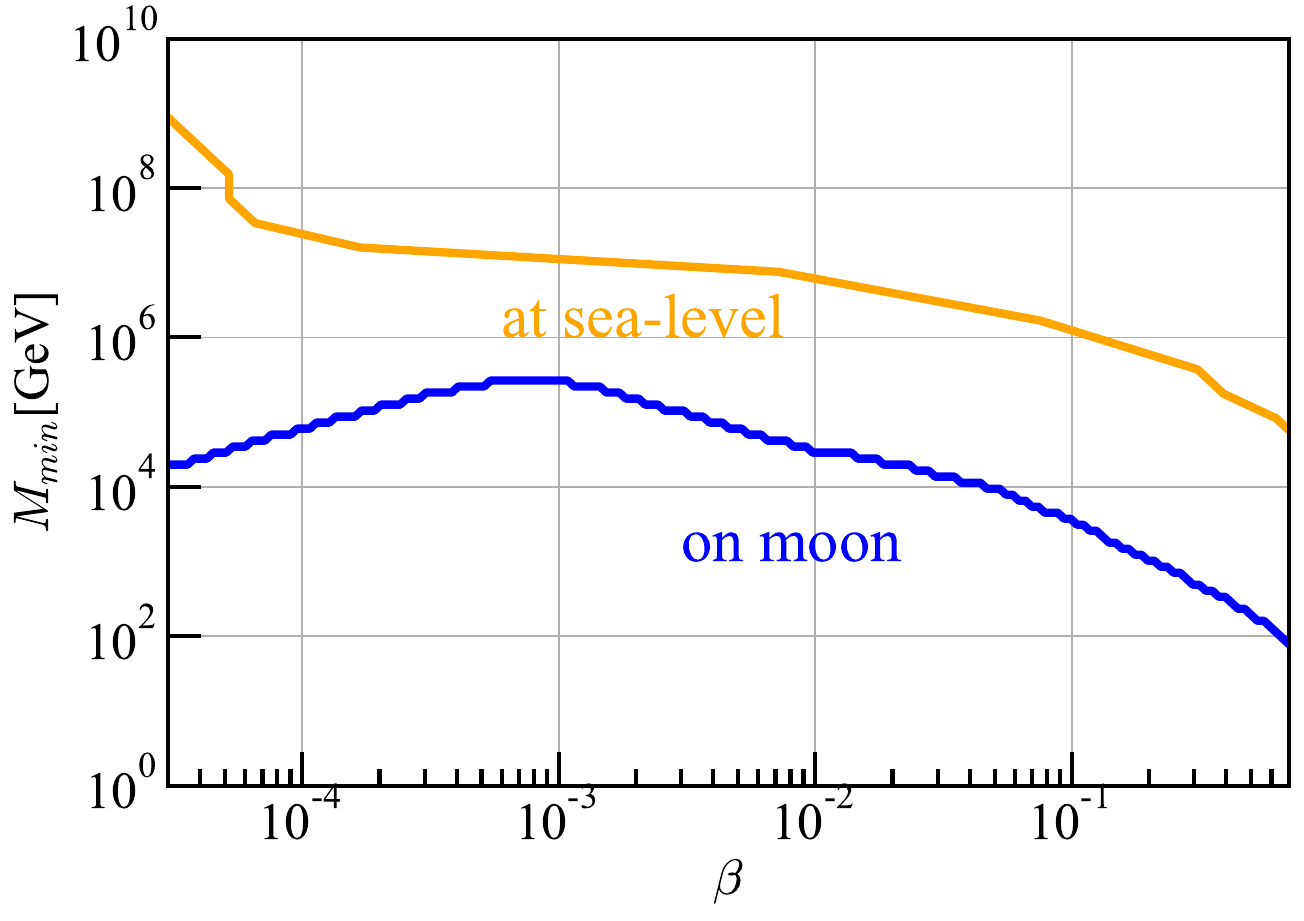}
    \caption{
    The minimal detectable masses of the cosmic MMs as a function of the MM $\beta$ for a moon-based detector and a terrestrial detector.
    }
    \label{fig:min_mass_vs_beta}
\end{figure}

\begin{figure}[htp]
    \centering
    \includegraphics[width=0.9\columnwidth]{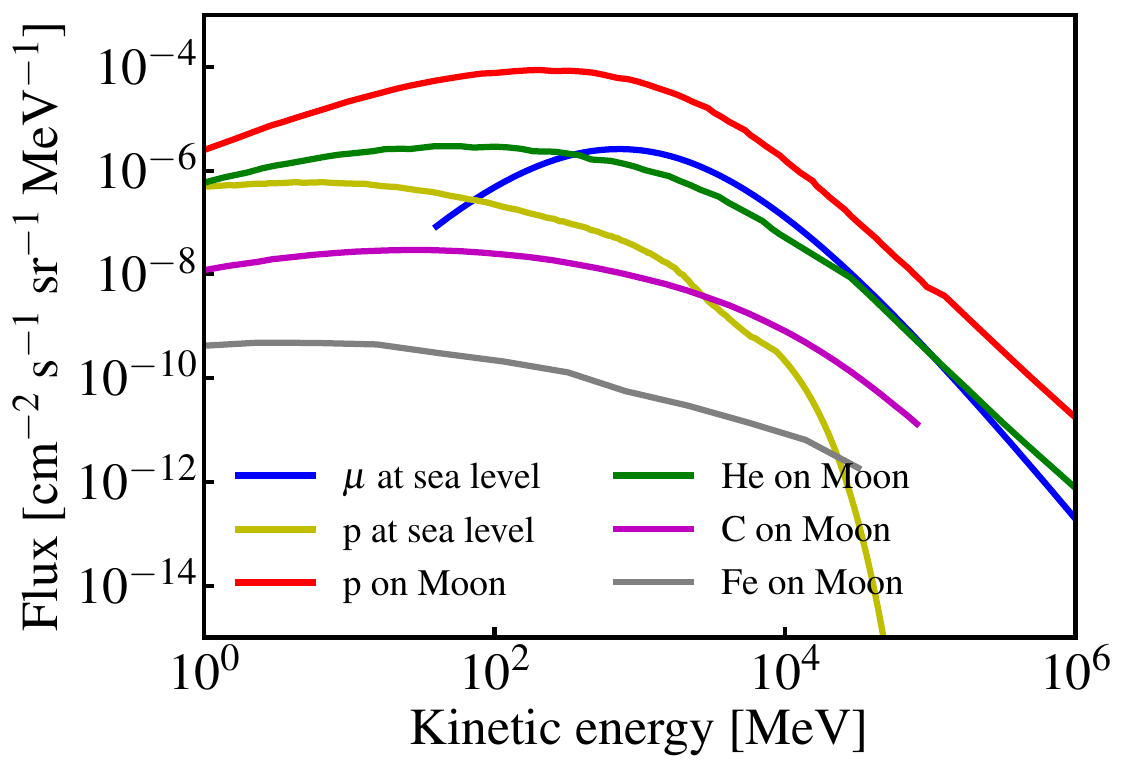}
    \caption{
    The fluxes of background particles as a function of their kinetic energies.
    The blue and yellow solid lines represent the muon and proton fluxes at sea level, which are calculated based on Bugaev/Reyna model~\cite{su2021comparison} and simulated by CRY algorithm~\cite{hagmann2007cosmic}, respectively.
    The red, green, magenta, and gray lines give the fluxes of proton, helium, carbon, and iron in space, taken from Ref.~\cite{waller2020simulated}.
    }
    \label{fig:particle_background_fluxes_moon}
\end{figure}

\begin{figure}[htp]
    \centering
    \includegraphics[width=0.9\columnwidth]{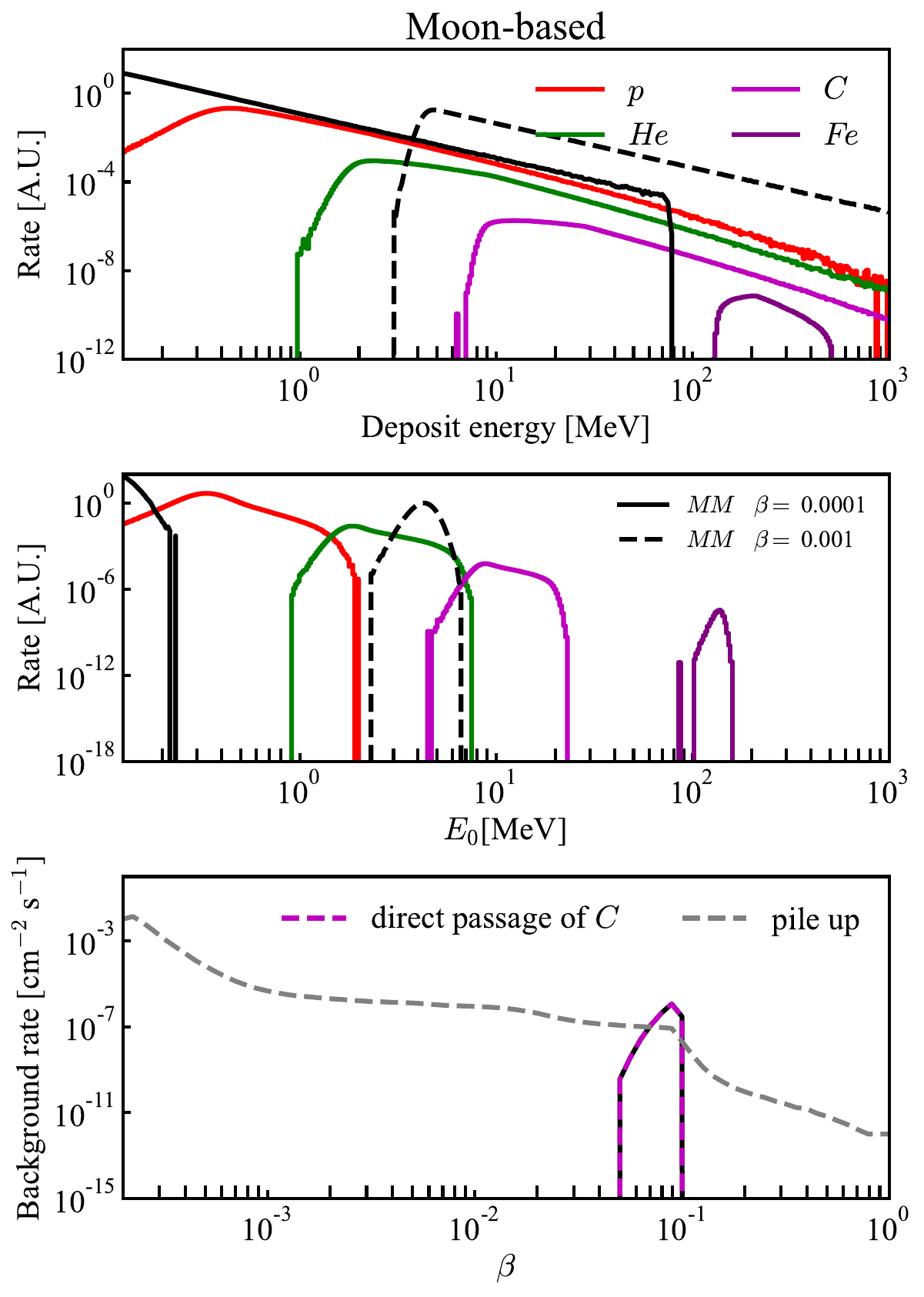}
    \caption{
    The background estimation for a moon-based detector.
    The simulated reconstructed energy distributions of the background particles and MMs with different speeds from all the angles are shown in the top panel.
    The middle panels show the $E_0$ distribution of background particles and MMs.
    The colored solid lines showing the contributions from different background particles, while the black solid and dashed lines show the distributions of MMs with $\beta$ of 0.0001 and 0.001, respectively.
    The lower panel shows the background rate after the energy and ToF requirements, including the direct passage component (colored dashed line) and pile up (grey line) component, as a function of assumed MM speed. }
    \label{fig:background_rate_vs_speed_moon}
\end{figure}

\begin{figure*}[htbp]
    \centering    \includegraphics[width=0.95\textwidth]{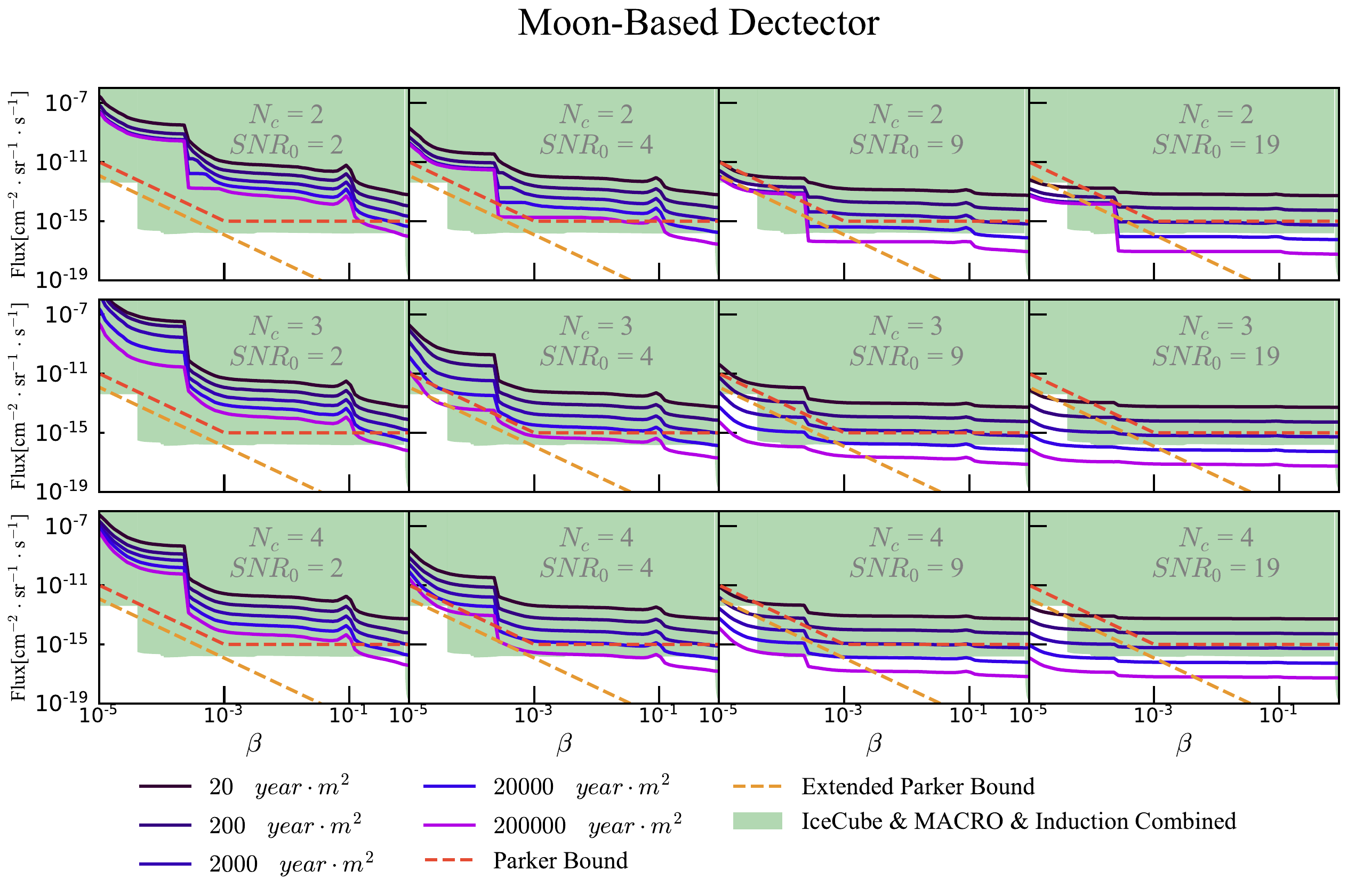}
    \caption{
    The projected sensitivity $\mathcal{Q}$ for a moon-based detector. 
    Each sub-figure in the table shows the projected sensitivity as a function of speed of MMs.
    The colored solid lines represent the sensitivities under different exposure as illustrated in the legends.
    The green region shows the flux constraints that have been given by MARCO~\cite{macro2002final}, IceCube~\cite{aartsen2016searches} and superconductivity induction experiments~\cite{giacomelli1984magnetic}. 
    Each row shows the sensitivity with different SNR$_0$ assumption and each column gives the sensitivities with different $N_c$.
    }
    \label{fig:space detector}
\end{figure*}

A moon-based deployment of the SCEP detector is of particular scientific interest, not only for the cosmic MM searches but also for the detection of other exotic particle phenomena.
The magnetometer of the SCEP detector can be specialized to search for the pseudo-magnetic fields that could be induced by the presence of exotic particles~\cite{jiang2021search}.
A super-long baseline detector network, with magnetometers located on both the Earth and the Moon, can potentially increase the sensitivity for the detection of the topological defect dark matters, such as domain walls~\cite{lohe1979soliton, gani2014kink, gani2015kink}. 
Similar Earth-based detector network concepts have been proposed and implemented in previous works~\cite{afach2021search}.
For MM searches specifically, a moon-based detector can extend the mass range of the search, as the Moon has no atmosphere and less intervening matter compared to a terrestrial detector.
This can lower the detectable MM mass threshold by approximately 2 to 4 orders of magnitude.
Fig.~\ref{fig:min_mass_vs_beta} shows the minimal detectable MM mass as a function of MM speed for both a moon-based and a terrestrial sea-level detector.
In the calculation of the minimal detectable MM mass for a moon-based detector, it is assumed that the MM must be able to penetrate a 5-cm thick PS layer to be registered by the detector.

However, the particle background environment on the lunar surface is more severe compared to the terrestrial environment. On the Moon, muons are no longer the dominant source of background, due to the absence of the atmosphere.
In the deep space environment surrounding the Moon, high-energy protons and helium nuclei emerge as the prevailing sources of the background particles, as shown by  AMS~\cite{aguilar2021periodicities, aguilar2022properties}, DAMPE~\cite{dampe2019measurement, dampe2022detection}, and CALET~\cite{marrocchesi2019measurement,brogi2022measurement} experiments. 
Assuming the negligible influence of the Earth's magnetic field on the lunar surface, it is conservatively assumed that the fluxes of protons, helium nuclei, carbon nuclei, and iron nuclei are approximately isotropic.
The fluxes of these background particle species as a function of their kinetic energies are taken from Ref.~\cite{waller2020simulated}, and are also shown in Fig.~\ref{fig:particle_background_fluxes_moon}.

The estimated particle background for a moon-based detector is shown in Fig.~\ref{fig:background_rate_vs_speed_moon}. 
The calculation is based on the methods described in Subsection~\ref{subsec:ionization_background}. 
The dominant background is due to the cosmic protons, helium nuclei, carbon nuclei, and iron nuclei. 
These high-energy particles cover the entire energy deposition spectrum up to the GeV range, overlapping with the expected energy distributions for the MM of different speeds.
As a result, the particle background for a moon-based detector is approximately 4 orders of magnitude higher than that of a sea-level terrestrial detector.
The sensitivity of a moon-based MM detector is given in Fig.~\ref{fig:space detector}, based on the calculations outlined in Subsection~\ref{subsec:sensitivity}. 
Due to the high particle background in the MM speed range greater than 2.5$\times$10$^{-4}$ light speed, a higher trigger threshold of the induction signal is required to achieve a ``background-free'' search.
This, in turn, lowers the overall acceptance of the detector.
Consequently, the sensitivity of a moon-based MM detector is not as competitive as that of a sea-level terrestrial detector. 
The required SNR$_0$ and N$_c$ are raised to be larger than 7.6 and 2, respectively, to overcome the elevated particle background environment on the lunar surface, with an assumed exposure of 20000 year$\cdot$m$^2$.


\end{document}